\documentclass[a4paper,fleqn]{cas-dc}

\usepackage[numbers]{natbib}
\usepackage{subcaption}
\usepackage{amsmath}
\usepackage{setspace}
\def\tsc#1{\csdef{#1}{\textsc{\lowercase{#1}}\xspace}}
\tsc{WGM}
\tsc{QE}
\tsc{EP}
\tsc{PMS}
\tsc{BEC}
\tsc{DE}

\begin{document}

\let\WriteBookmarks\relax
\def\floatpagepagefraction{1}
\def\textpagefraction{.001}

  \addtolength\abovedisplayskip{-0.5\baselineskip}%
  \addtolength\belowdisplayskip{-0.5\baselineskip}%

\title [mode = title]{Holographic Transmitarray Antenna with linear Polarization in X band}
\tnotemark[0]

\author[1]{Mahdi Salehi}[type=,
                        auid=000,bioid=000,
                        prefix=,
                        role=,
                        orcid=0000-0003-1182-1608]

\address[1]{Iran University of Science and Technology, School of Electrical Engineering, Tehran, 16846-13114, Iran}

\author[1]{Homayoon Oraizi}[]
\cormark[1]

\cortext[cor2]{Corresponding author.}
\ead{h_oraizi@iust.ac.ir}

\begin{abstract}
For the first time, we present the design and demonstration of holographic transmitarray antennas (TAs) based on the susceptance (reactance) distribution in this paper. According to the holographic theory, the amplitudes and phases of electromagnetic waves can be recorded on a surface, and then they can be reconstructed independently. This concept is used to design single-beam and multi-beam linearly polarized holographic TAs without using any time-consuming optimization algorithms. Initially, an impedance surface is analyzed for both transmission and reflection modes. As there are differences between the susceptance (reactance) distribution of these modes, a new approach (different from reflectarray designs) is proposed to apply the holographic technique to transmitarray designs. Then, interferograms are described based on the scalar transmission susceptance distribution according to the number and direction of the radiation beams. Subsequently, a transmission metasurface of dimensions equal to $0.26\lambda_0$ is hired to design holographic TAs at 12 GHz. Several holograms are designed using the unit cell to verify the proposed method. Finally, as a proof of concept, a linearly polarized circular aperture wideband holographic transmitarray antenna with a radius of 13.3 cm is manufactured and tested. The antenna achieves 12.5\% (11.4-12.9 GHz) 1-dB gain bandwidth and 23.8 dB maximum gain, leading to 21.46\% aperture efficiency. Furthermore, the antenna achieves 95.94\% simulated radiation efficiency mainly due to using subwavelength elements, which becomes possible by applying the holographic technique.
\end{abstract}

\begin{keywords}
Holographic Technique\sep
Transmitarray Antenna \sep
Metasurface\sep
Transmission and Reflection Impedance surface\sep
\sep Wideband Antenna
\end{keywords}

\maketitle

\section{Introduction}

Transmitarrays have attracted much interest in recent years because of their inherent advantages such as high gain, high radiation efficiency, lightweight, low profile, and simple manufacturing procedure, making the antenna the right candidate for radar applications(e.g., Surveillance radars, Synthetic Aperture radars, and Marine radars), point-to-point telecommunications, and broadcasting systems. A transmitarray is illuminated through space feeding, so compared with the phased array, it does not experience the insertion loss of the feeding network at high frequency. Besides, as the transmitarray is illuminated directly from the back of the aperture, it avoids blockage losses expected with the reflectarray antennas.
\par The idea of holographic antenna was first published in 1969 by P.checcacci, V. Russo and A.M. Scheggi \cite{a1}.  Technological developments in PCB and CNC machining have encouraged researchers to improve the idea. Up to now, several holographic antennas (holographic leaky wave antennas and holographic reflectarray antennas) have been designed. In Ref. 2, two holographic reflectarray antennas with linear and circular polarizations are proposed. The linearly polarized antenna achieves 25\% aperture efficiency, and the circularly polarized antenna achieves 15.57\% aperture efficiency. In Ref. 3, a multi-beam linearly polarized holographic reflectarray antenna is proposed. It achieves 22.2\% 1-dB gain bandwidth and 53.09\% aperture efficiency for dual beam radiating. In Ref. 4, a circular polarized holographic reflectarray antenna is proposed. The antenna can produce multiple beams with independent circular polarizations. It achieves 19.23\% 1-dB gain bandwidth and 45.8\% aperture efficiency. In Ref. 5, a holographic reflectarray with the capability of producing shaped electromagnetic waves is proposed. The antenna can produce multiple beams with arbitrary shapes. In Ref. 6 to 10, holographic leaky wave antennas are proposed. In Ref.11, a horn-fed holographic transmitarray antenna is proposed to generate high spatial resolution and power-efficient 3D fields.  The research uses the dyadic Green's function (DGF) as the propagation kernel of fields, directly linking sources and fields to obtain the phase distribution describing the hologram. The research proposes a dual-polarized unit cell consisting of two cross shape metallic layers sandwiched by a dielectric layer with the periodicity of $\lambda_{0}/2$ at 20 GHz and uses it to realize the transmission phases describing the hologram. Two 3D patterns of the fields describing cartoon images of "Panda" and "Monkey" are generated in x, and y polarizations, respectively. The measured results show that the proposed method is accurate and can generate high-resolution 3D fields in Microwave Therapy, Volumetric Printing, superoscillation, and two-photon microfabrication. In Ref.12, a holographic lens antenna is proposed to convert the spherical waves generated by two dipole feeds to vortex waves of OAM numbers ($l_{1}=+1$, $l_{2}=-1$). The proposed lens antenna generates the above OAM modes with an isolation level of better than 20 dB, developing two reliable communication links at 60 GHz. The research proposes a unit cell capable of changing the input wave polarization from the x-polarized to the y-polarized wave and vice versa with the periodicity of $0.33\lambda_{0}$ to realize the transmission phase distribution describing the hologram.
\par According to the authors' knowledge, most of the researches into the holographic antenna concept have been limited to reflectarray and leaky wave antennas. However, there are some works that designed holographic transmitarrays based on the transmission phase distribution \cite{a11,a12}. Therefore, a detailed study of designing holographic transmitarrays based on the reactance (susceptance) distribution is presented here for the first time. We present the design of horn-fed linearly polarized holograms for generating single and dual linearly polarized pencil beams at the X band. As a linearly polarized communication channel has greater capacity than a circularly polarized channel and does not experience the 3-dB loss expected with the circularly polarized channel, linear polarization is chosen for the hologram designs.The holographic technique makes the TA antenna design process easy and fast. It allows a reduction in the number of antenna layers by limiting the required transmission phase from $360^\circ$ to $180^\circ$ (-$90^\circ$: $90^\circ$). It improves the TA performance, such as radiation efficiency (due to using subwavelength elements), operational bandwidth, and cross-pol level, comparing with the only-phase approach.
\par This paper is organized as follows: Section 2 analyzes an impedance surface for both reflection and transmission modes for the first time. Section 3 proposes an equation to obtain the desired susceptance (reactance) distribution to radiate the object wave. It explains how to apply the holographic technique to tranmitarray designs, based on the reactance(susceptance) distribution. Section 4 proposed a unit cell and presents the design and simulation of the unit cell. Section 5 provides several holograms to validate the design procedure proposed in section 3.  All holograms are simulated using the CST Studio Suite software. As a proof of concept, a prototype is fabricated and tested, obtaining close agreement with simulation results. Section 6 discusses the measurement and simulation results. Finally, section 7 offers the conclusion of this study.

\section{Transmission and Reflection Impedance surfaces}

\begin{figure}[!t]
\centering
	\centering\includegraphics[width =0.45 \textwidth]{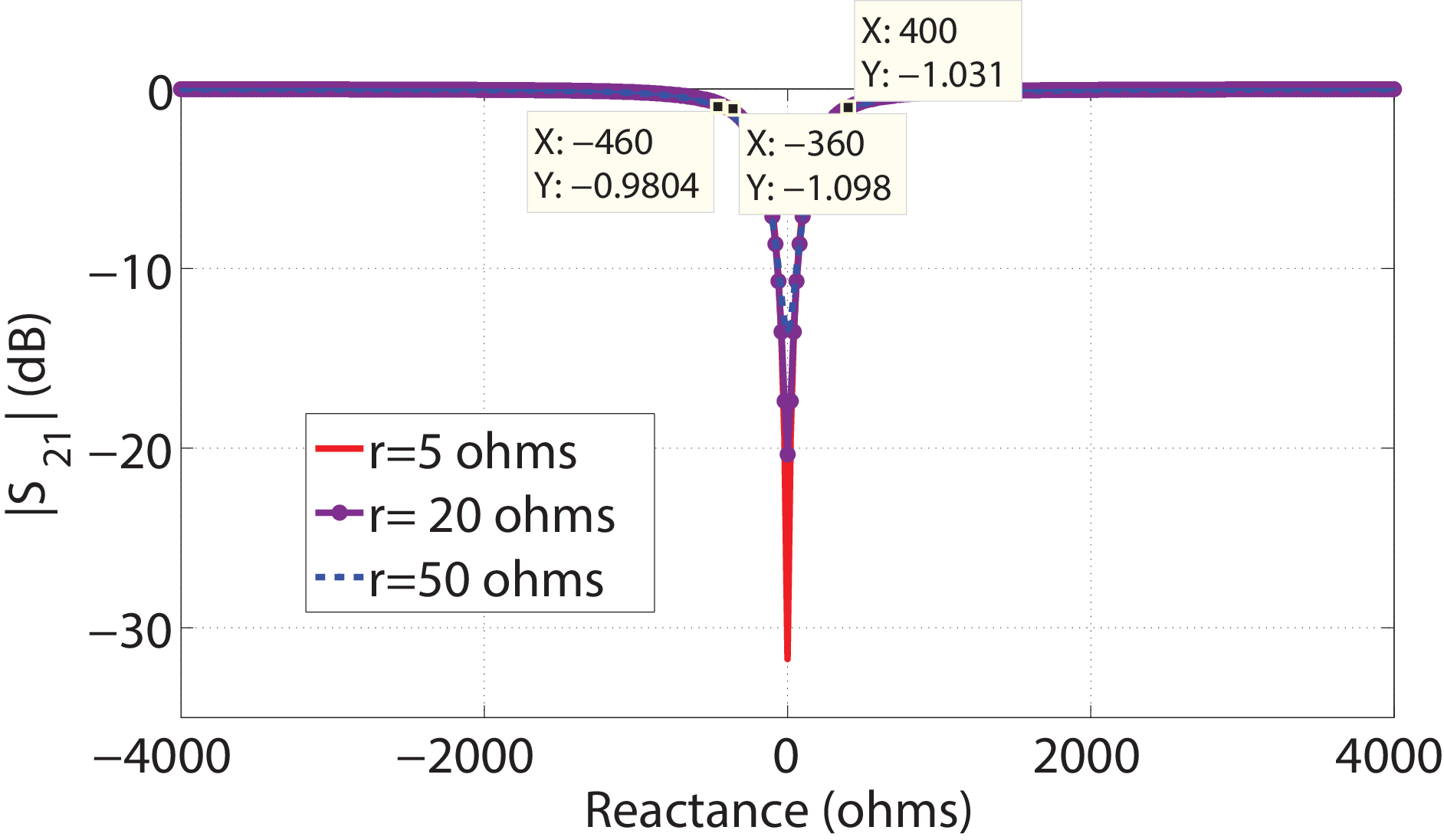}
\caption{ The $|S_{21}|$ curves as a function of the reactance values for the resistance values of 5, 20, and 50 ohms are shown. Note that the sheet is a square patch with a length of 4 mm, and the simulation frequency is 15 GHz. In addition, the impedance of the Floquet ports are 50 ohms.}
\label{fig:fig1}

\end{figure}

An impedance surface reflects and transmits waves simultaneously with different coefficients. In order to study the coefficients in more depth, an impedance sheet with variable reactance values and three different resistance values of 5, 20, and 50 ohms is simulated, as a unit cell, using the ANSYS HFSS software. Because increasing the resistance values generate losses, low resistance values are selected. Master/Slave boundaries are forced on the four sides of the unit cell to simulate an infinite array of sheets, and the impedance values of the Floquet ports are 50 ohms. The transmission magnitudes are shown in Fig.\ref{fig:fig1}. The figure shows that the $|S_{21}|$ curve jumps over -1 dB level for reactance values greater than +360 ohms and less than -360 ohms, greater than +400 ohms and less than -400 ohms, and greater than +460 and less than -460 ohms when the resistance value equals, 5, 20, and 50 ohms, respectively. In other words, depending on the resistance value, the sheet starts changing its reaction to the electromagnetic waves from reflection to transmission when the reactance value increases from a specific value and decreases from another specific value (e.g., ±360,±400, and ±460 are -1 dB  limits when the sheet resistance equals 5, 20, and 50, respectively.) This is a critical difference between RA unit cells and TA unit cells, which necessitates applying a method different from what has been proposed in Ref.2 to 5.

\begin{figure}[!t]
		\centering\includegraphics[width =0.47 \textwidth]{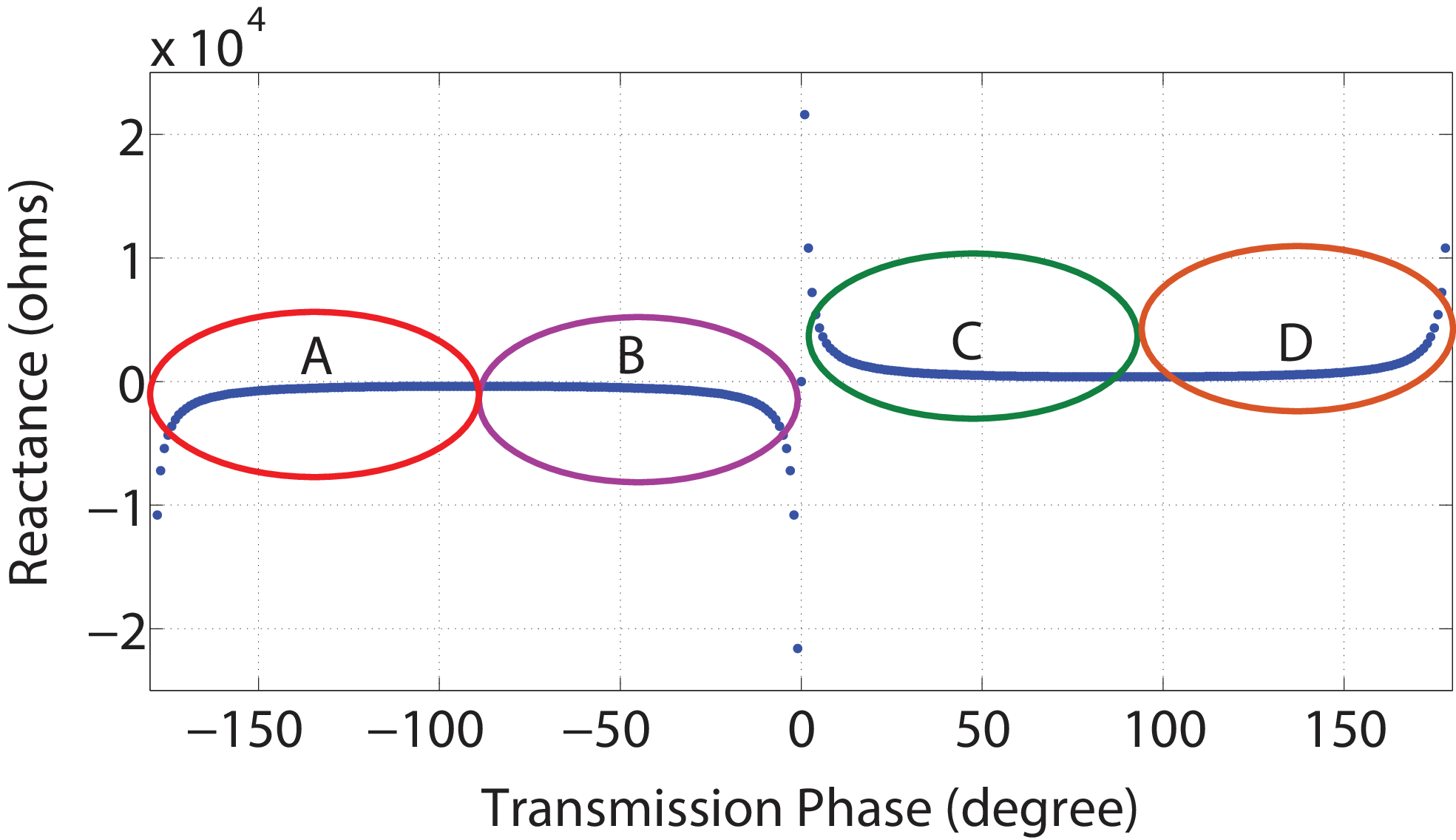}
\caption{The reactance values in terms of the transmission phases under the following assumptions: lossless unit cell $|S_{11}|=0$, and $|S_{12}|=1$. Apparently, A=B , and C=D.}
\label{fig:fig2}

\end{figure}
\begin{figure}[!t]
		\centering\includegraphics[width =0.47 \textwidth]{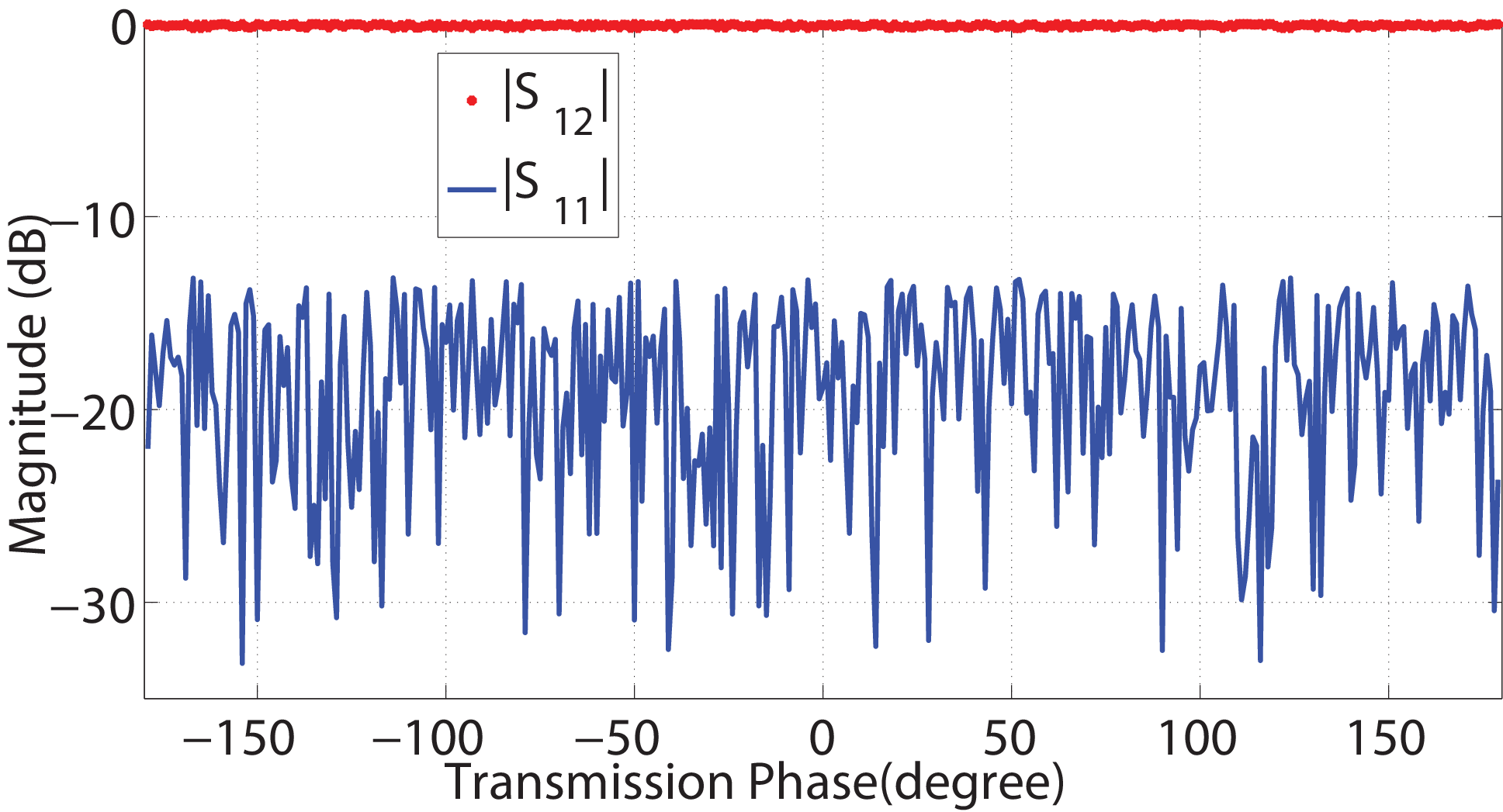}
\caption{ The $|S_{11}|$ and $|S_{12}|$ values in terms of the transmission phases under the lossless condition are shown. The $|S_{11}|$ magnitudes are random values between 0.02 (-33.97 dB) and 0.22 (-13.15 dB) generated by MATLAB, and the $|S_{12}|$ magnitudes are obtained from Eq.(1).}
\label{fig:fig3}
\end{figure}

\begin{figure}[!t]
\centering\includegraphics[width =0.47 \textwidth]{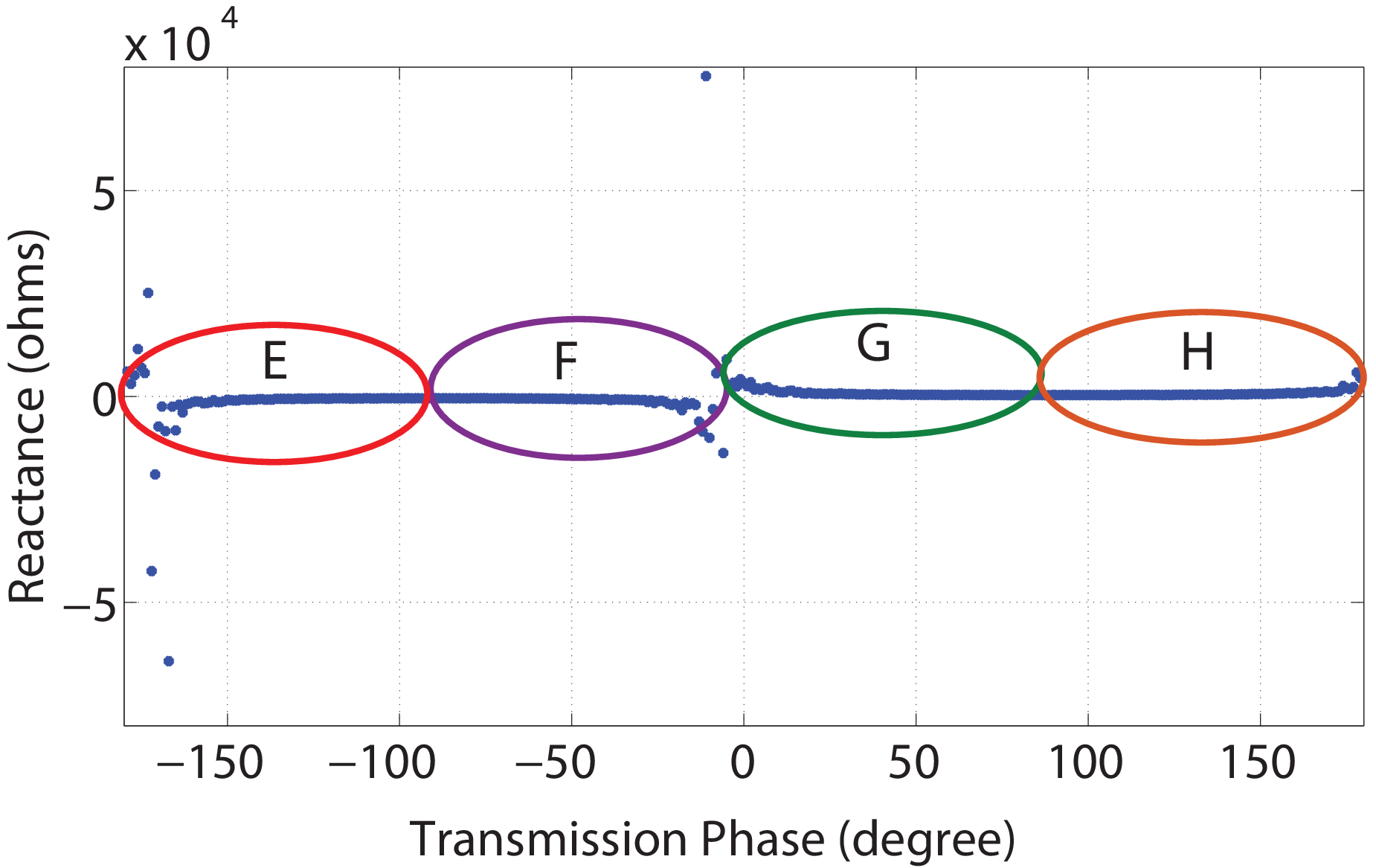}
\caption{The reactance values in terms of the transmission phases under the lossless conditions are shown. The $|S_{11}|$ magnitudes are random values between 0.02 (-33.97 dB) and 0.22 (-13.15 dB) generated by MATLAB, and the $|S_{12}|$ values are obtained from Eq.(1). In addition, the reflection phase $\alpha_{1}$ is obtained from Eq.(2). Apparently, E=F , and G=H.}
\label{fig:fig4}
\end{figure}
\par The $Z_{12}$ coefficients of a two-port network as functions of the scattering parameters are obtained from Eq.(1) \cite{a13}. A transmission metasurface is a lossy two-port network. However, lossless unit cells are considered to simplify the calculations. Lossless conditions under the assumption of  $S_{12}=S_{21}$ are expressed in Eq.(2) and  Eq.(3), where $S_{11}=|S_{11}|e^{(j\alpha_{1})}$, $S_{22}=|S_{22}|e^{(j\alpha_{2})}$, and $S_{21}=S_{12}=|S_{21}|e^{(j\beta)}$  \cite{a13}. A lossless   unit cell under the assumptions of $S_{11}=S_{22}=0$, $S_{12}=S_{21}$, and ($\beta)=[-180^\circ, +180^\circ]$ is examined, although it is an impractical mathematical model. This model is studied as a guide to predict the reactance (susceptance) distribution of a real unit cell. By substituting the above assumptions to Eq.(1) and performing a few calculations, Eq.(4) is obtained. The reactance curve versus the transmission phase ($\beta$) is shown in Fig.\ref{fig:fig2}. As seen in Fig.\ref{fig:fig2}, the reactance values repeat when the transmission phase varies from $0^\circ$ to $90^\circ$ and $90^\circ$ to $180^\circ$ (C=D). Following the same routine, the reactance values repeat when the transmission phase varies from $-180^\circ$ to $-90^\circ$ and $-90^\circ$ to $0^\circ$ (A=B). For more clarification, assume that $|S_{11}|$ varies between 0.02 (-33.97 dB) and 0.22 (-13.15 dB). As it is impractical to assign the same $|S_{11}|$ values to different                  unit-cells, MATLAB is used to generate random $|S_{11}|$ values to apply a general model. Therefore, the $|S_{21}|$ values can be obtained from Eq.(2). Besides, the transmission phase $(\beta)$ range is assumed to be $[-180^\circ, +180^\circ]$, and the range of the reflection phase ($\alpha_{1}$) is obtained from Eq.(3). The transmission and reflection magnitudes are shown in Fig.\ref{fig:fig3}. By substituting the aforementioned assumptions in Eq.(1), the reactance values are obtained and shown in Fig.\ref{fig:fig4}. As seen in Fig.\ref{fig:fig4}, the reactance values repeat when the transmission phase varies from $0^\circ$ to $90^\circ$ and $90^\circ$ to $180^\circ$ (G=H). Following the same routine, the reactance values repeat when the transmission phase varies from $-180^\circ$ to $-90^\circ$ and $-90^\circ$ to $0^\circ$ (E=F). As seen in  Fig.\ref{fig:fig3} and Fig.\ref{fig:fig4}, it is required for the unit cell to provide only $180^\circ$ phase shifts ($-90^\circ$ to $90^\circ$) to cover all possible reactance (susceptance) values.
 However, if the goal is to design a TA based on the only-phase technique, the applied unit cell should provide $360^\circ$ phase range to compensate for the phase delays at different positions on the antenna aperture so that the antenna can convert the incident field phase front from spherical to the plane phase front. The holographic technique reduces the number of layers and air gaps in the cell's structure by reducing the required phase range. Furthermore, it allows using subwavelength metasurfaces, which improves antenna bandwidth and radiation efficiency.

 \begin{equation}
Z_{12}=\frac {2\eta_{0}S_{21}}{(1-S_{11})(1-S_{22})-S_{12}S_{21}}\label{eq2}
\end{equation}
 \begin{equation}
|S_{11}|^{2}+|S_{21}|^{2}=1\label{eq1}
\end{equation}
 \begin{equation}
S_{11}S^{*}_{21}+S_{22}S^{*}_{21}=0\label{eq2}
\end{equation}
 \begin{equation}
Z_{12}=j\eta_{0}\frac{1}{\sin(\beta)}\label{eq2}
\end{equation}

\section{Holographic Transmitarray Description}

The following steps summarize the procedure of designing a holographic transmitarray: 1- 	designing the mathematical hologram and sampling it with a sampling rate associated with the cell periodicity to obtain the interferogram 2- 	Implementing the impedance (admittance) description of the hologram by using quasi-periodic unit cells. The mathematical hologram is obtained from Eq.(5) \cite{a4}, so the problem is with the practical realization of the hologram. In Ref. 10, the utilization of artificial impedance surfaces with modulated dimensions is proposed to solve this problem. According to Eq.(5), the holographic interference pattern is purely amplitude while it contains the phase information of the interfered waves \cite{a14}.

\begin{multline}
Z(x,y)|Y(x,y)=j\left(X+M\Re\left(\psi_{obj}\psi_{ref}^*\right)\right) \label{eq1}
\end{multline}
\setlength{\belowdisplayskip}{0pt}
\par The parameters X and M are arbitrary average impedance (admittance) value and arbitrary modulation depth, respectively. These values must be assigned so that the impedance (admittance) distribution of the hologram can be correctly implemented by the unit cell. After sampling the mathematical hologram (described by Eq.(5)), the usable impedance (admittance) distribution is obtained, making the realization of the hologram possible in the form of an array of quasi-periodic unit cells with variable patch size. The cell's patch size is determined to meet the local impedance (admittance) value of the hologram surface. Also, its dimensions (sampling rate) are determined to satisfy the Nyquist rate.

 \par
\begin{figure}[!t]
\centering
	\centering\includegraphics[width =0.45 \textwidth]{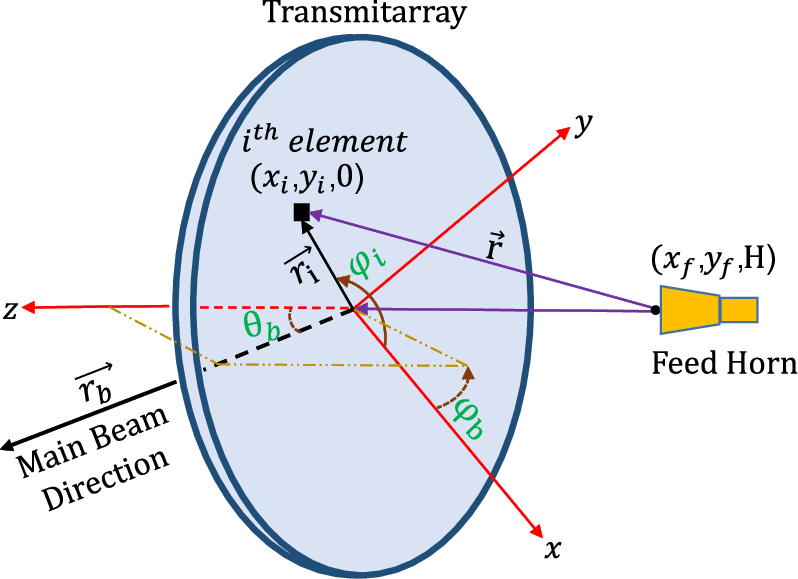}
\caption{ The schematic of the feed horn, transmitarray, and coordinate system}
\label{fig:fig5}
 \end{figure}

 For a linear space feed holographic transmitarray antenna, the reference wave is a spherical wave emitted by the horn antenna, and the object wave is a plane wave with a directed beam at ($\theta_b$,$\phi_b$), as defined in Eq.(6) and Eq.(7), where $\vec{r}_i=x_i\hat{x}+y_i\hat{y}$ and  $\vec{r}_b=(\sin(\theta_b)\cos(\phi_b)\hat{x}+\sin(\theta_b)\cos(\phi_b)\hat{y}+\cos(\theta_b)\hat{z})$, as illustrated in Fig.\ref{fig:fig5}. By substituting Eq.(6) and Eq.(7) in Eq.(5), the mathematical hologram is obtained from Eq.(8). Note that ($x_f$,$y_f$,H) designates for the feed position.
\setlength{\abovedisplayskip}{0pt}
\begin{multline}
\psi_{obj}(x_i,y_i)=e^{-jk_0(\vec{r}_i.\vec{r}_b)}=\\
e^{-jk_0(x_i\sin(\theta_b)\cos(\phi_b)+y_i\sin(\theta_b)\sin(\phi_b))}\label{eq2}
\end{multline}
\begin{equation}
\psi_{ref}(x_i,y_i)=e^{-jk_0r}=e^{-jk_0\sqrt{(y_i-y_f)^2+(x_i-x_f)^2+H^2}}\label{eq3}
\end{equation}
\begin{multline}
Z(x,y)|Y(x,y)=j\Bigl(X+M\Bigl(\cos\Bigl(k_0(r-\\
x_i\sin(\theta_b)\cos(\phi_b)-y_i\sin(\theta_b)\sin(\phi_b))\Bigr)\Bigr)\Bigr)\label{eq4}
\end{multline}
\setlength{\belowdisplayskip}{0pt}

 \par  According to Fig.\ref{fig:fig1}, a transmission artificial impedance surface cannot generate reactance values between ± 360,± 400, and ± 460 ohms while experiencing $|S_{21}|$ better than -1 dB when the resistance value equals 5, 20, and 50 ohms, respectively. Eq.(5) is a modulated cosine function and generates values between X±M. Depending on X and M values, X+M and X-M can be lower or upper bands of the reactance (susceptance)range, which may have positive or negative signs. If both bands have positive signs, Eq.(5) generates only-inductive reactance values, and if they have negative signs, Eq.(5) generates only-capacitive reactance values. As it is required to construct holograms by using inductive and capacitive unit-cells ( create resonances), X+M and X-M must have opposite signs. In this case, Eq.(5) generates reactance values between ±360, ±400, and ±460 ohms. It may also generate values greater than +360 ohms and less than -360 ohms, greater than +400 ohms and less than -400 ohms, and greater than +460 ohms and less than -460 ohms. For example, when X=-100 and M=800, the reactance values lie between -900 and +700 ohms, including between ±360, ±400, and ±480 ohms. Another example, when  X=-15 and M=105, the reactance values lie between -120 and 90 ohms, as subsets of between ±360, ±400, and ±460 ohms. Therefore, Eq.(5) cannot be used to implement a holographic TA, while it can be used to implement holographic RA, similar to the ones proposed in Ref. 2 and Ref. 5. We dub this impedance zone "holographic reflectarray antenna zone (HRAZ)" or "holographic transmitarray forbidden zone (HTAFZ)." Here, a solution is proposed to avoid HTAFZ. First, Eq.(5) is converted to two equations by creating two conditions. Second, two arbitrary constants are added to these equations. Because X is the arbitrary average impedance (admittance) value and does not contain the amplitude and phase information of the object waves, we are allowed to change it. The new impedance (admittance) description of the hologram is given by Eq.(9):
where A and B are the two arbitrary constants. By adequately determining these values, the hologram can pass the HTAFZ.
The holographic technique can be applied to generate more than one object wave. In this case, the hologram records the information (phase and amplitude) of all object waves. The superposition of the object waves is expressed in Eq.(10), where $\psi^{i}_{obj}$ is the $i^{th}$ object wave.

\begin{equation}
\begin{cases}
Z_{new}|Y_{new}=Z_{old}|Y_{old}+A & if (Z_{old}|Y_{old}$<0$)\\
Z_{new}|Y_{new}=Z_{old}|Y_{old}+B & if (Z_{old}|Y_{old}$>0$)
\end{cases}
\end{equation}

\begin{equation}
\psi^{total}_{obj}=\left( \sum_{i=1}^{i=n}{\psi^{i}_{obj}}\right) \label{eq6}\\
\end{equation}

\section{Unit Cell Design}

\begin{figure}[!t]
\centering
	\begin{subfigure}[b]{0.24\textwidth}
		\includegraphics[width = \textwidth]{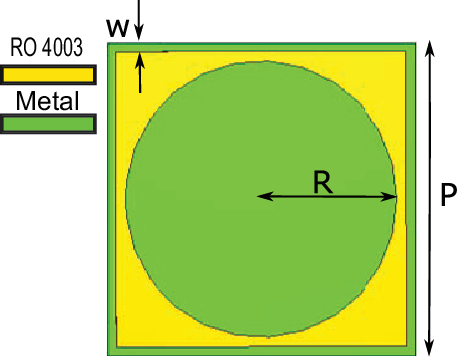}
		\caption{}
		\label{fig:fig6(a)}
	\end{subfigure}
	\begin{subfigure}[b]{0.19\textwidth}
		\includegraphics[width = \textwidth]{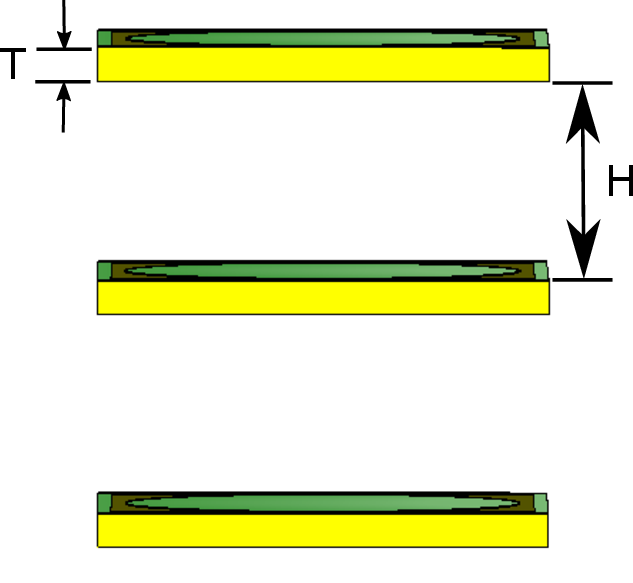}
		\caption{}
		\label{fig:fig6(b)}
	\end{subfigure}
	\begin{subfigure}[b]{0.24\textwidth}
		\includegraphics[width = \textwidth]{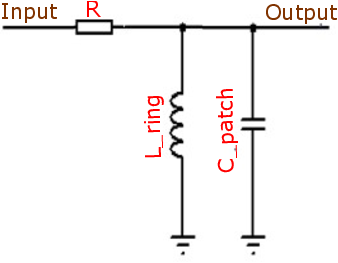}
		\caption{}
		\label{fig:fig6(c)}
	\end{subfigure}

\caption{The triple layer unit-cell layout  (a) top view, (b) side view, and (c) the equivalent circuit of one layer unit-cell. The cell dimensions are as follows: H=6 mm, T=0.508 mm,w=0.2 mm, R (0.1 mm to 2.9 mm), and P=6.6 mm. Note that the green parts represent metal, and the yellow parts represent dielectric.}
\end{figure}

\begin{figure*}[!t]
\centering
	\begin{subfigure}[b]{0.33\textwidth}
		\includegraphics[width = \textwidth]{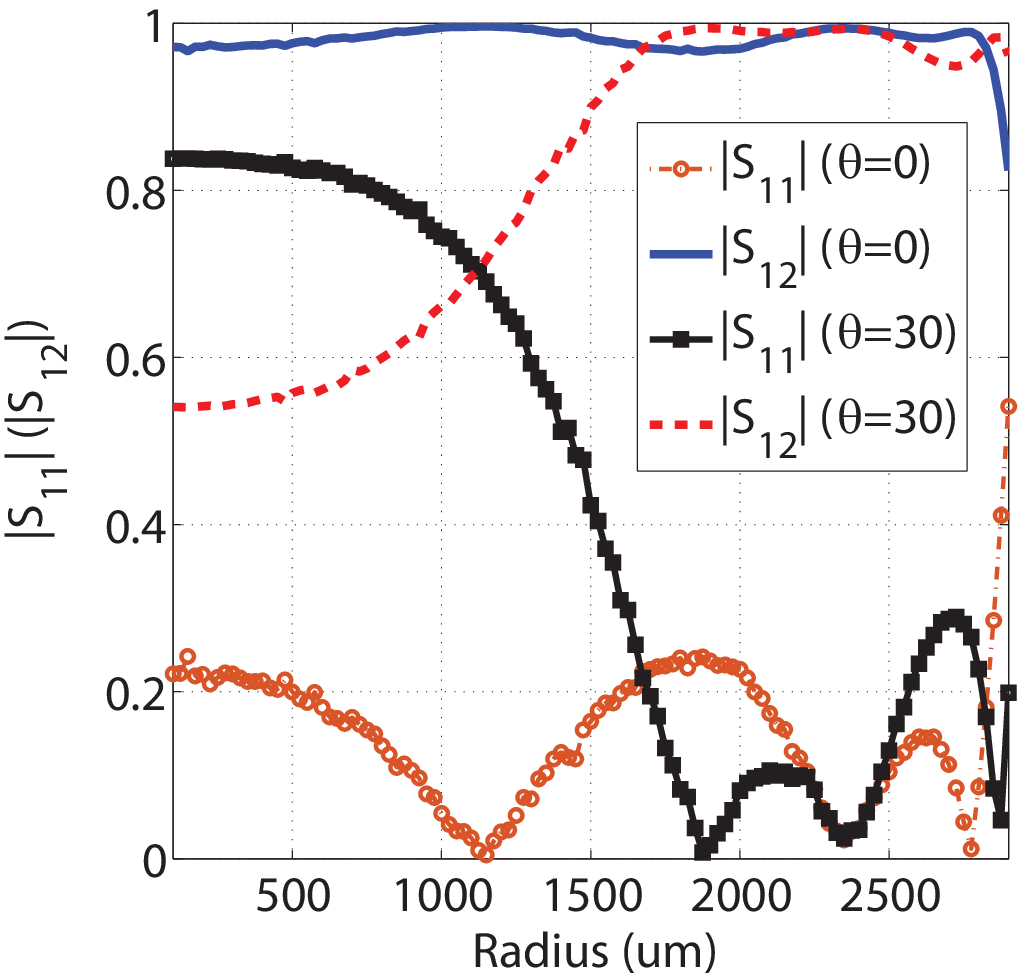}
		\caption{}
		\label{fig:fig7(a)}
	\end{subfigure}
	\begin{subfigure}[b]{0.33\textwidth}
		\includegraphics[width = \textwidth]{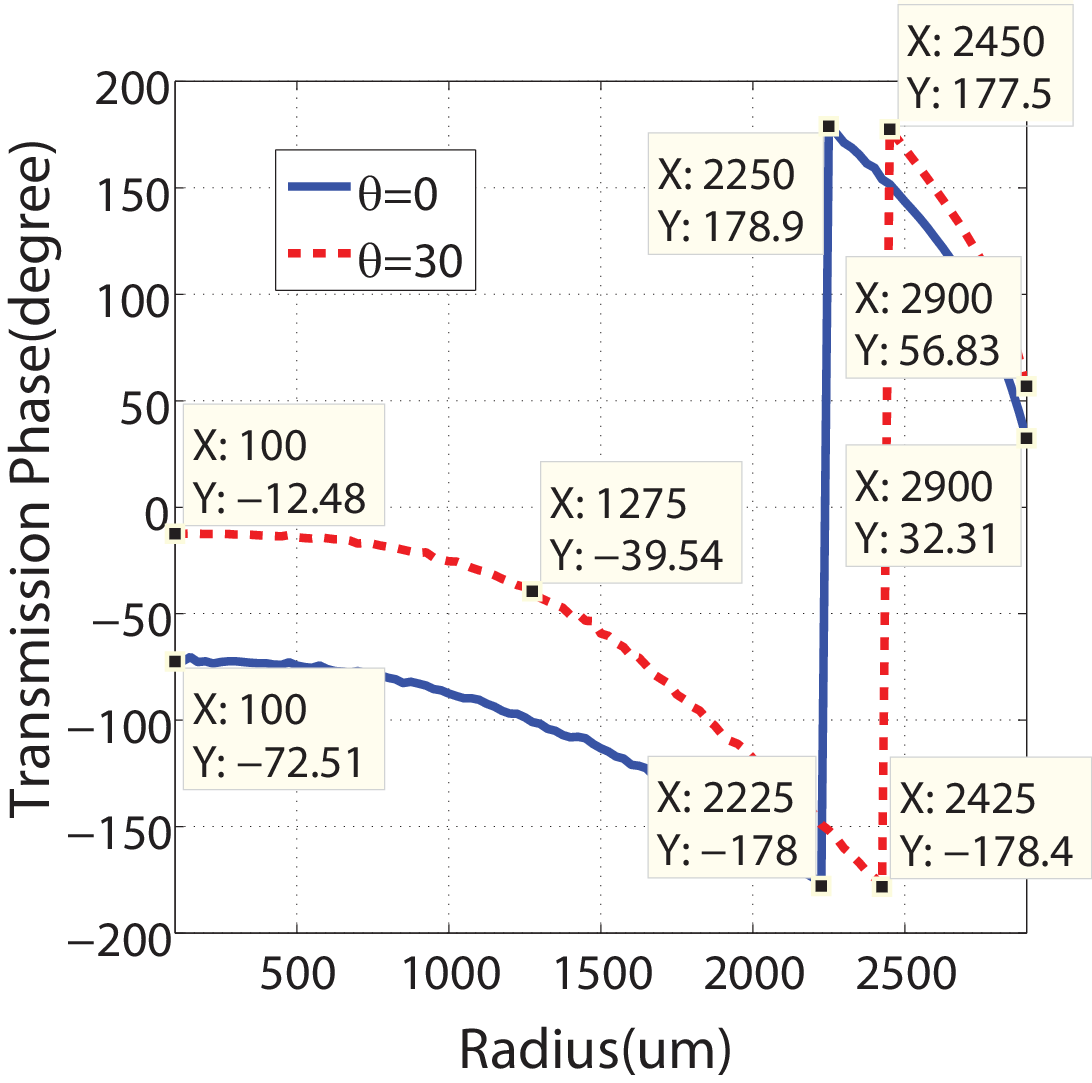}
		\caption{}
		\label{fig:fig7(b)}
	\end{subfigure}
	\begin{subfigure}[b]{0.33\textwidth}
		\includegraphics[width = \textwidth]{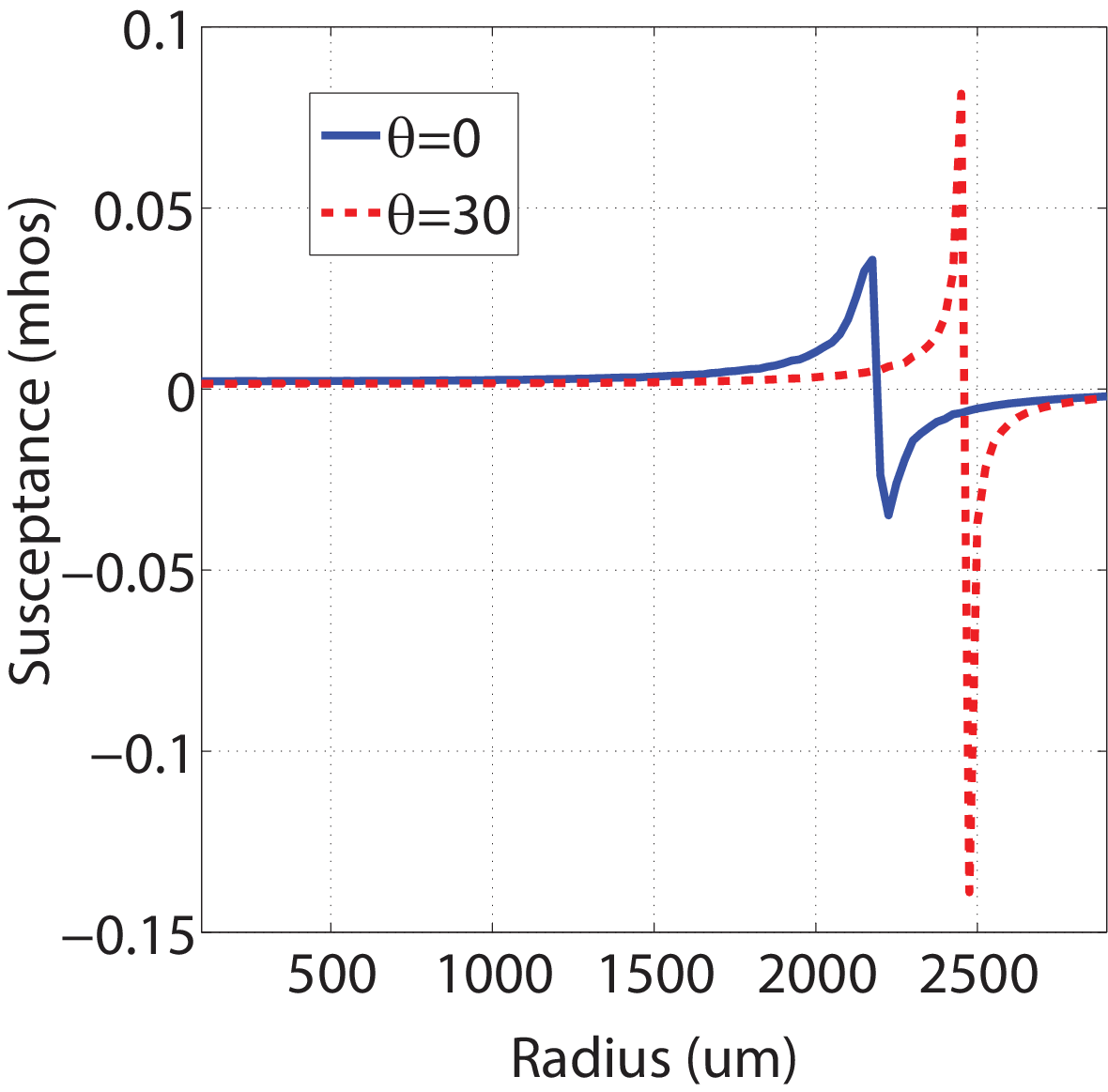}
		\caption{}
		\label{fig:fig7(c)}
	\end{subfigure}
	\begin{subfigure}[b]{0.35\textwidth}
		\includegraphics[width = \textwidth]{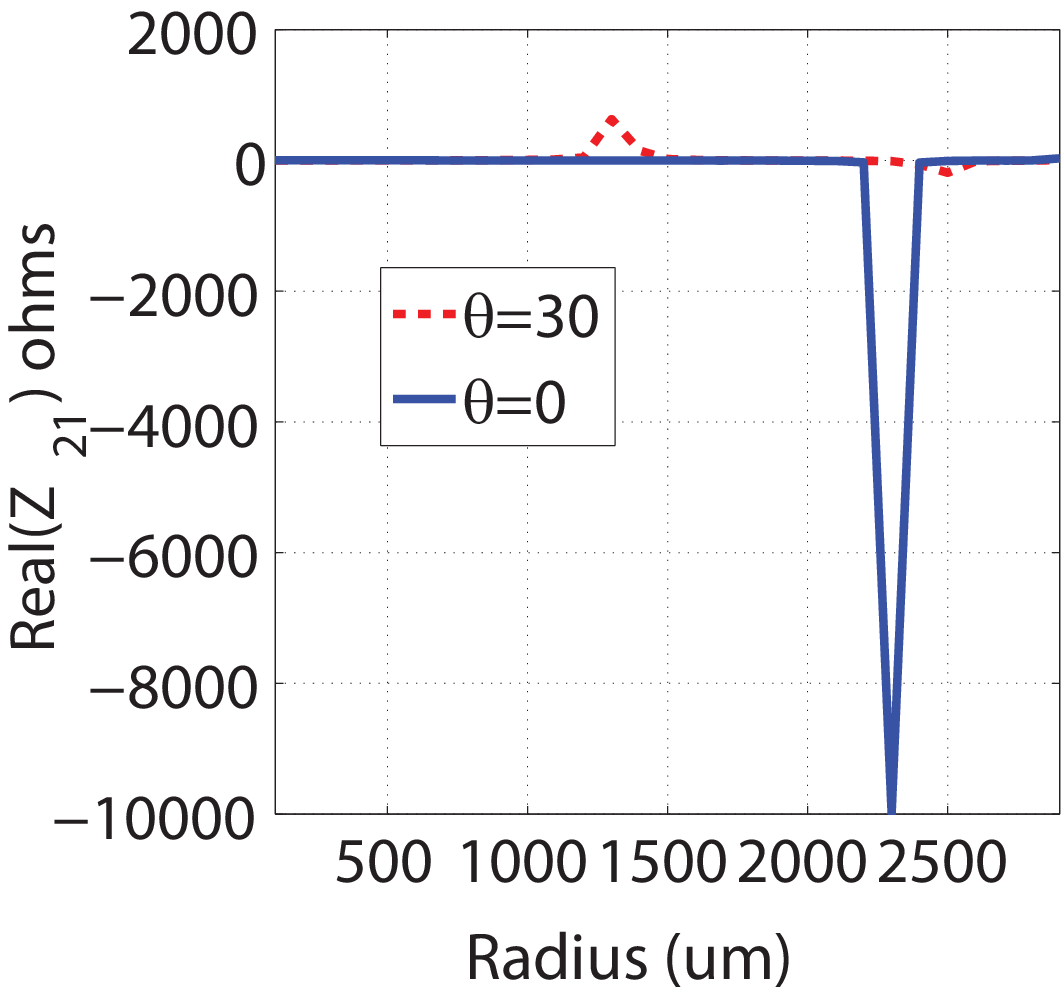}
		\caption{}
		\label{fig:fig7(d)}
	\end{subfigure}

\caption{(a) Simulated $|S_{11}|$ and $|S_{11}|$, (b) Simulated transmission phases},(c) Simulated susceptance values, and (d)Simulated real($Z_{21}$) values. Note that the circular patch radius R belongs on the x-axis of all graphs, and the simulation is done for incident angles of $\theta=0^\circ$, and $\theta=30^\circ$ at 12 GHz.
\end{figure*}

As seen in Fig.\ref{fig:fig6(a)} and Fig.\ref{fig:fig6(b)}, a unit cell consisting of three identical metal layers of square ring elements and circular patch elements printed on three dielectric layers made by Rogers 4003C $(\epsilon_r=3.55, tan\gamma=0.0027, T= 0.508 mm)$ is employed. There is an air gap between every two layers with the height of H=$0.24\lambda_0$=6 mm, which is set to generate almost $90^\circ$ phase shifts while considering the standard length of plastic spacers. According to the simulation results, if the cell periodicity is less than $0.26\lambda_0$, the unit cell can not provide the transmission phase range required to generate all possible reactance (susceptance) values by varying the radius length of the circular patch. Therefore, one layer must be added to the unit cell structure to add extra resonance and generate the remaining transmission phase shifts. However, adding one air gap and one layer(dielectric and metal) increases the insertion loss and the cost. Thus, the cell periodicity is decided to be P=$0.26\lambda_0$=6.6 mm. Eq.(9) determines the susceptance (reactance) value of each pixel with the center point (x,y) on the mathematical hologram aperture. If the space between the center of each pixel and its adjacent pixels equals $\lambda/1000$, the hologram implementation will be impracticable because of the impossibility of designing unit cells with such periodicity. Therefore, it is essential to locally sample the mathematical hologram and then realize the susceptance(reactance) samples by unit cells. According to the Nyquist theory, the sampling rate has to be greater than half of the period of the reactance (susceptance) variations in the mathematical reactance (susceptance) distribution to avoid undersampling \cite{a3}. The value assigned to the cell periodicity (P=$0.26\lambda_0$=6.6 mm) meets the Nyquist rate. Fig.\ref{fig:fig6(c)} shows the equivalent circuit of every layer. The circular patch (shunt capacitor) creates a low-pass response, while the square ring (shunt inductor) creates a high-pass response. These two elements create a band-pass response. The resistor represents losses due to the substrate thickness. The square ring width is decided to be W=0.2 mm to keep the resonance frequency around 12 GHz by varying the circular patch radius length from 0.1 mm to 2.9 mm and, simultaneously, to generate a flat transmission magnitude.
The cell is simulated under the periodic boundary conditions using the CST Microwave Studio frequency-domain solver at 12 GHz. Simulation results are presented in Fig.\ref{fig:fig7(a)} up to Fig.\ref{fig:fig7(d)}. According to Fig.\ref{fig:fig7(a)}, transmission magnitudes better than -1 dB are obtained by varying the radius length from 0.1 mm to 2.85 mm under the normal incident angle. In addition, transmission magnitudes better than -1 dB are obtained by varying the radius length from 1.5 mm to 2.9 mm under the incident angle of $\theta=30^\circ$. As seen in Fig.\ref{fig:fig7(b)}, the unit cell provides -$178^\circ$ to -$72.5^\circ$ and $32.3^\circ$ to $178.9^\circ$ phase shifts when illuminated under the normal incident angle. Besides, it provides -$178.4^\circ$ to -$40^\circ$ and $56.8^\circ$ to $177.5^\circ$ phase shifts when illuminated under the incident angle of$(\theta=30^\circ)$. Although the unit cell can provide almost $250^\circ$ transmission phase shifts, only radius lengths that generate -$180^\circ$ to -$90^\circ$ and $90^\circ$ to $180^\circ$ phase shifts are taken into account.

\par
\begin{figure}[!t]
\centering
\includegraphics[width =0.48 \textwidth]{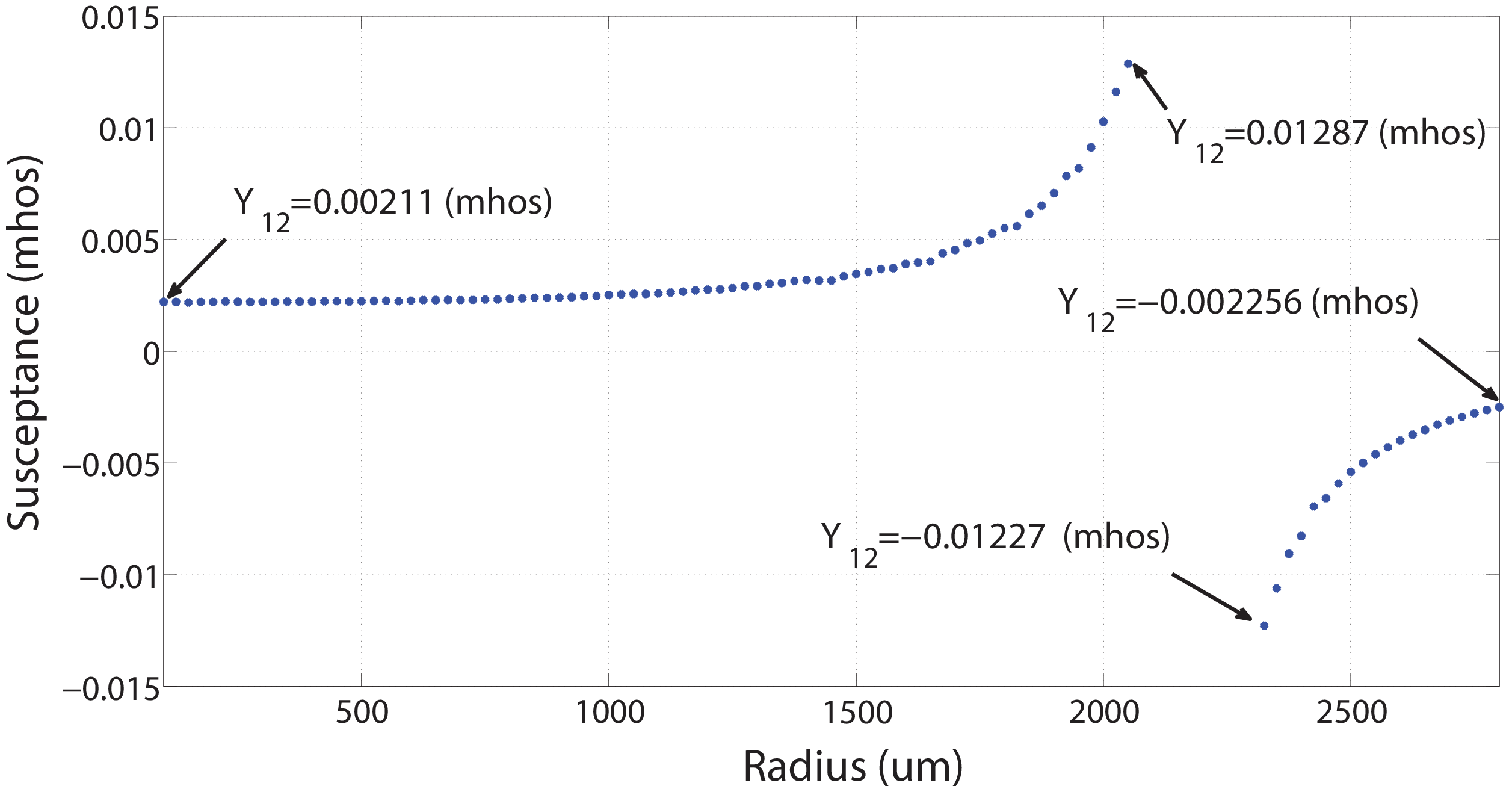}
\caption{ Simulated susceptance values in terms of the circular patch radius R under the normal incident angle at 12 GHz. The bounds are specified with arrows. The maximum susceptance value is 0.01287, and the minimum value equals -0.01227.}
\label{fig:fig8}
\end{figure}

 The transmission susceptance and resistance values are presented in Fig.\ref{fig:fig7(c)} and Fig.\ref{fig:fig7(d)}, respectively, for incident angles of $\theta=0^\circ$ and $\theta=30^\circ$. According to the figures, the susceptance curves have resonance points where the resistance values increase or decrease sharply, indicating some losses. Therefore, it is necessary to eliminate the resonance points from the available admittance range, as seen in Fig.\ref{fig:fig8}. Furthermore, Fig.\ref{fig:fig7(a)} up to Fig.\ref{fig:fig7(d)} show that transmission coefficients change considerably by varying the incident angle. These discrepancies increase the sidelobe and backlobe level of the final antenna. Thus, the unit cell is simulated under various incident angles from $0^\circ$ to $50^\circ$, with step $5^\circ$ to avoid the above disadvantages. It is worthwhile to mention that $|S_{12}|$ better than -1 dB is set as a criterion for all incident angles to avoid high backlobe level. Eq.(11) expresses the relation between $S_{21}$ and the Z parameters \cite{a13}. It is evident that $Z_{21}$ is not the only factor determining the $|S_{21}|$ value, and the sign of the resistance value alone cannot determine whether the power is generated or lost. Thus, although the resistance values are negative, the power is lost mainly by generating heat as the $|S_{21}|$ values are less than 0 dB.

\begin{equation}
S_{21}=\frac{2Z_{0}Z_{21}}{(Z_{11}+Z_{0})(Z_{22}+Z_{0})-Z_{12}Z_{21}}\label{eq11}
\end{equation}

\section{Hologram Design}

\begin{figure*}[!t]
\centering
	\begin{subfigure}[b]{0.245\textwidth}
		\includegraphics[width = \textwidth]{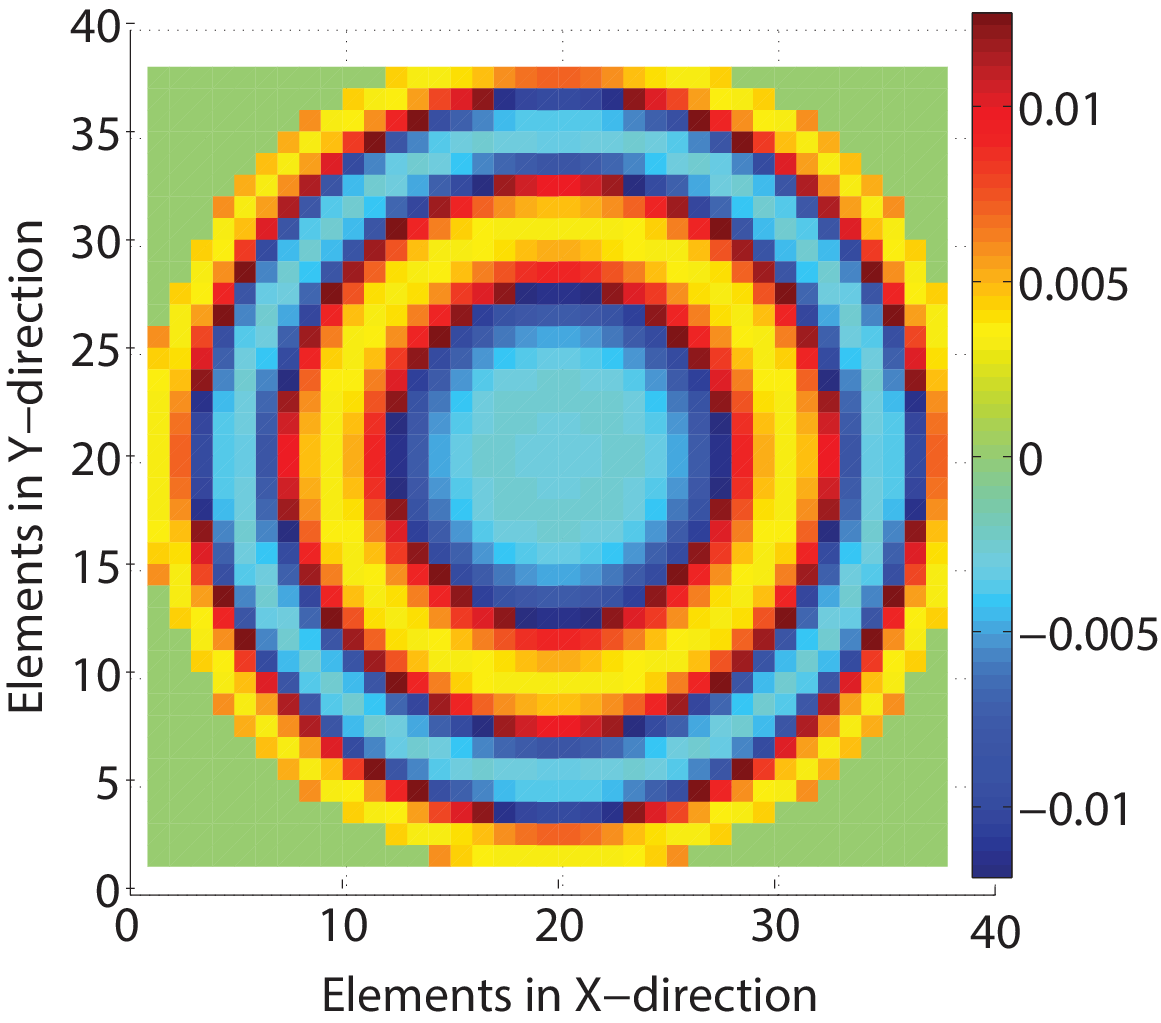}
		\caption{}
		\label{fig:fig9(a)}
	\end{subfigure}
	\begin{subfigure}[b]{0.245\textwidth}
		\includegraphics[width = \textwidth]{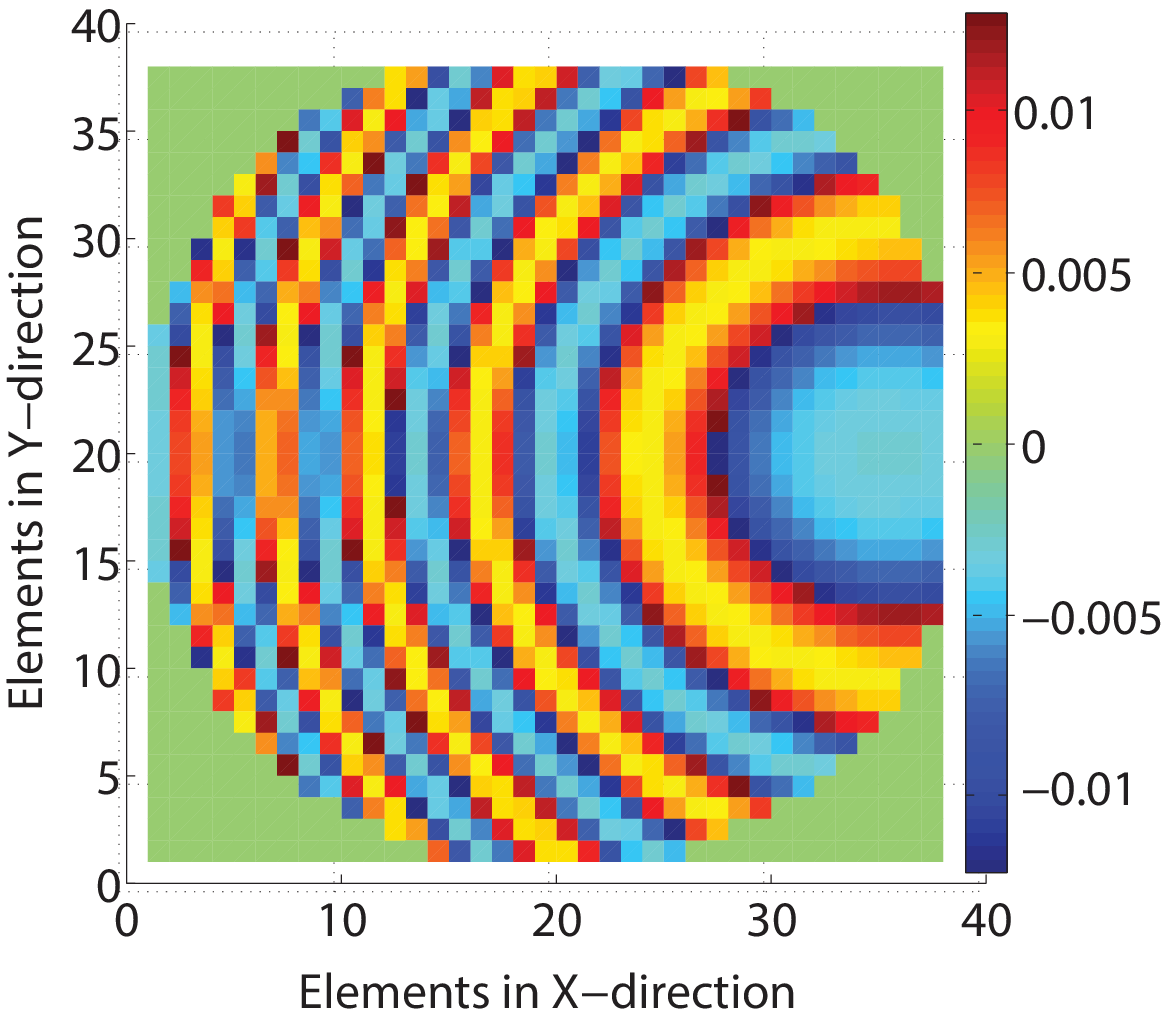}
		\caption{}
		\label{fig:fig9(b)}
	\end{subfigure}
	\begin{subfigure}[b]{0.245\textwidth}
		\includegraphics[width = \textwidth]{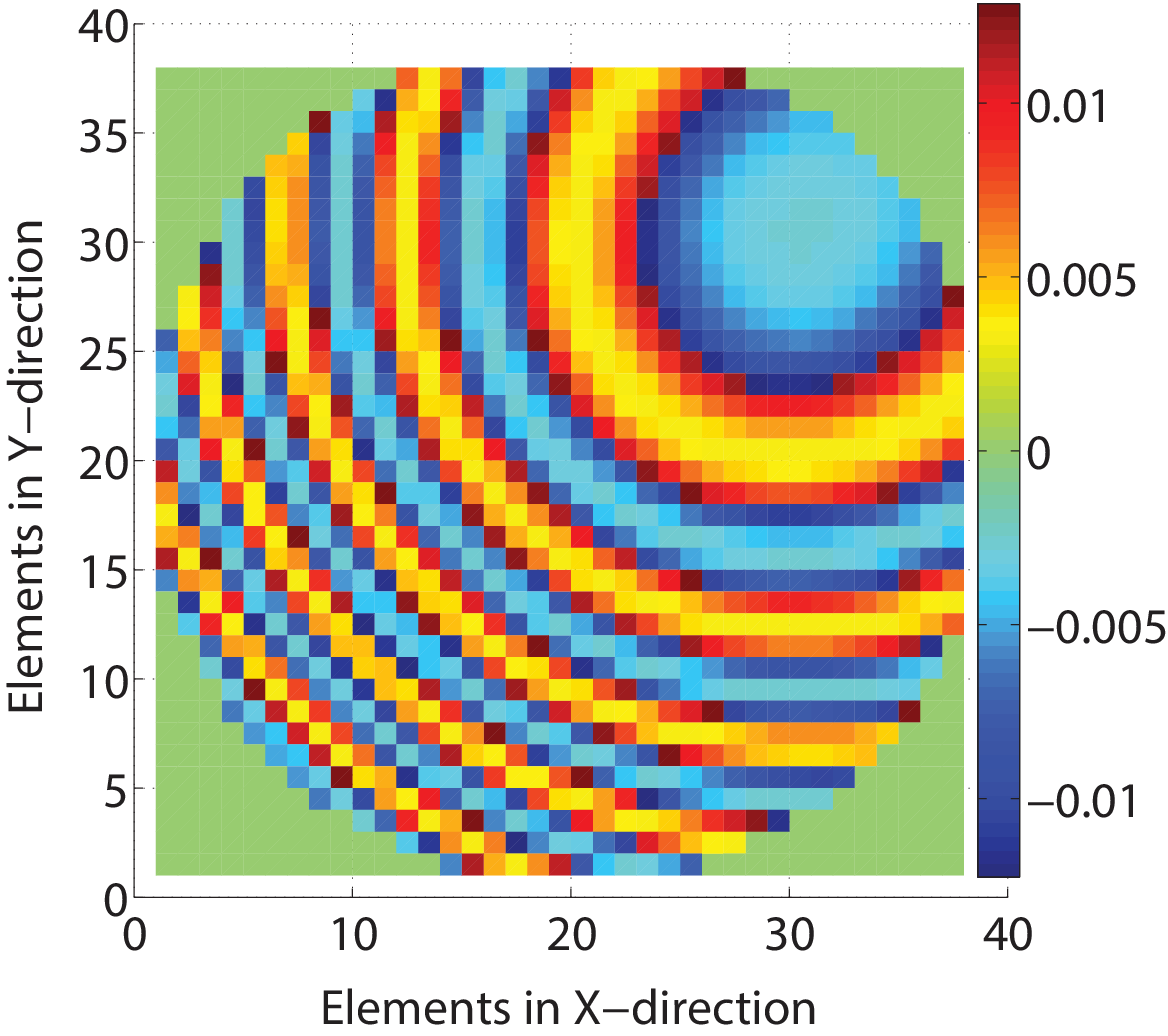}
		\caption{}
		\label{fig:fig9(c)}
	\end{subfigure}
\begin{subfigure}[b]{0.245\textwidth}
		\includegraphics[width = \textwidth]{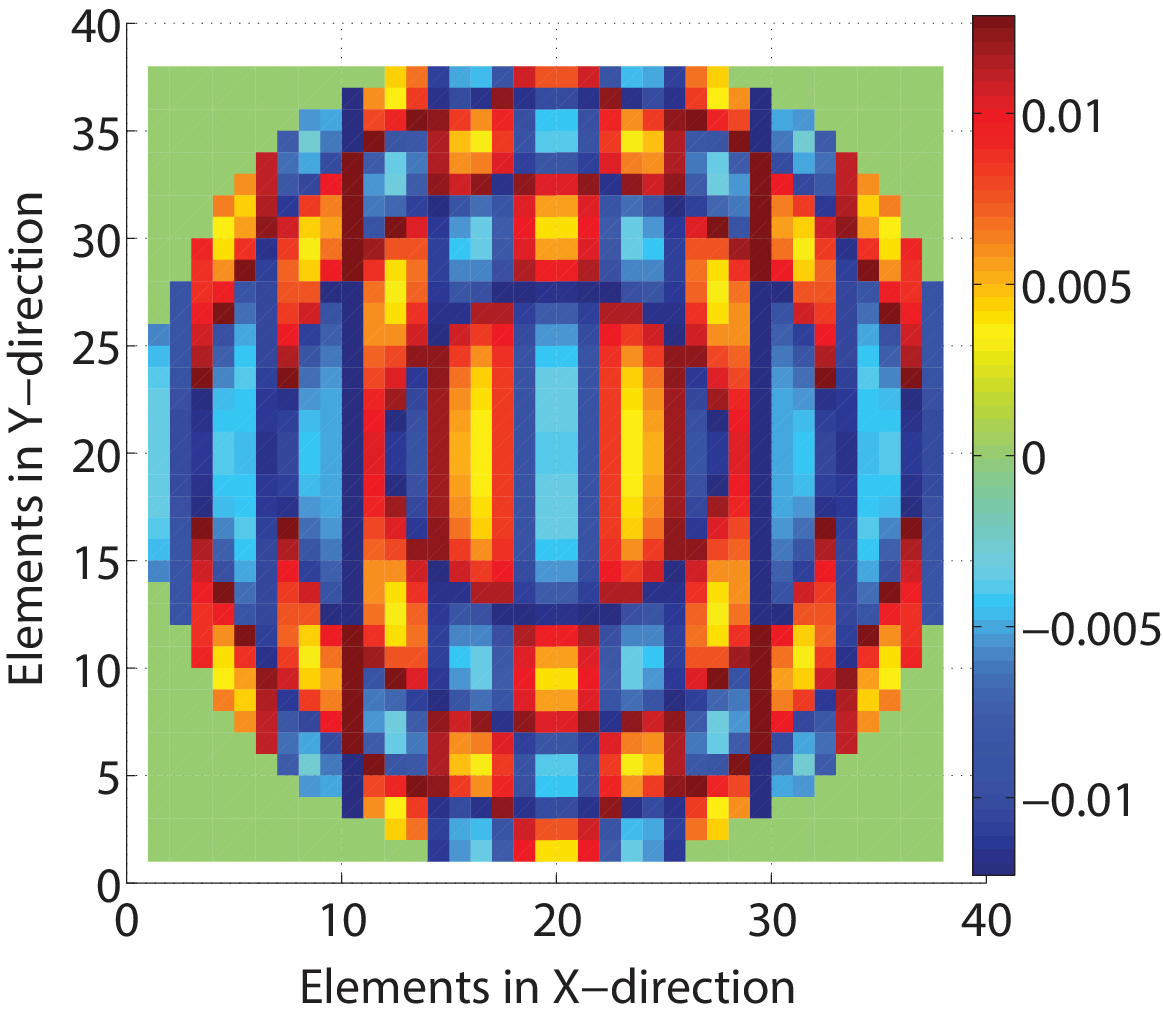}
		\caption{}
		\label{fig:fig9(d)}
	\end{subfigure}
\caption{Susceptance distribution of the holograms (a) Example 1  ($\theta=0^\circ,\phi=0^\circ$) (b) Example 2 ($ \theta=30^\circ,\phi=0^\circ$), (c) Example 3 ($\theta=30^\circ,\phi=45^\circ$), and (d) Example 4 ($ \theta=30^\circ,\phi_1=0^\circ(\phi_2=180^\circ)$)}
\end{figure*}
\begin{figure*}[!t]
\centering
	\begin{subfigure}[b]{0.32\textwidth}
		\includegraphics[width = \textwidth]{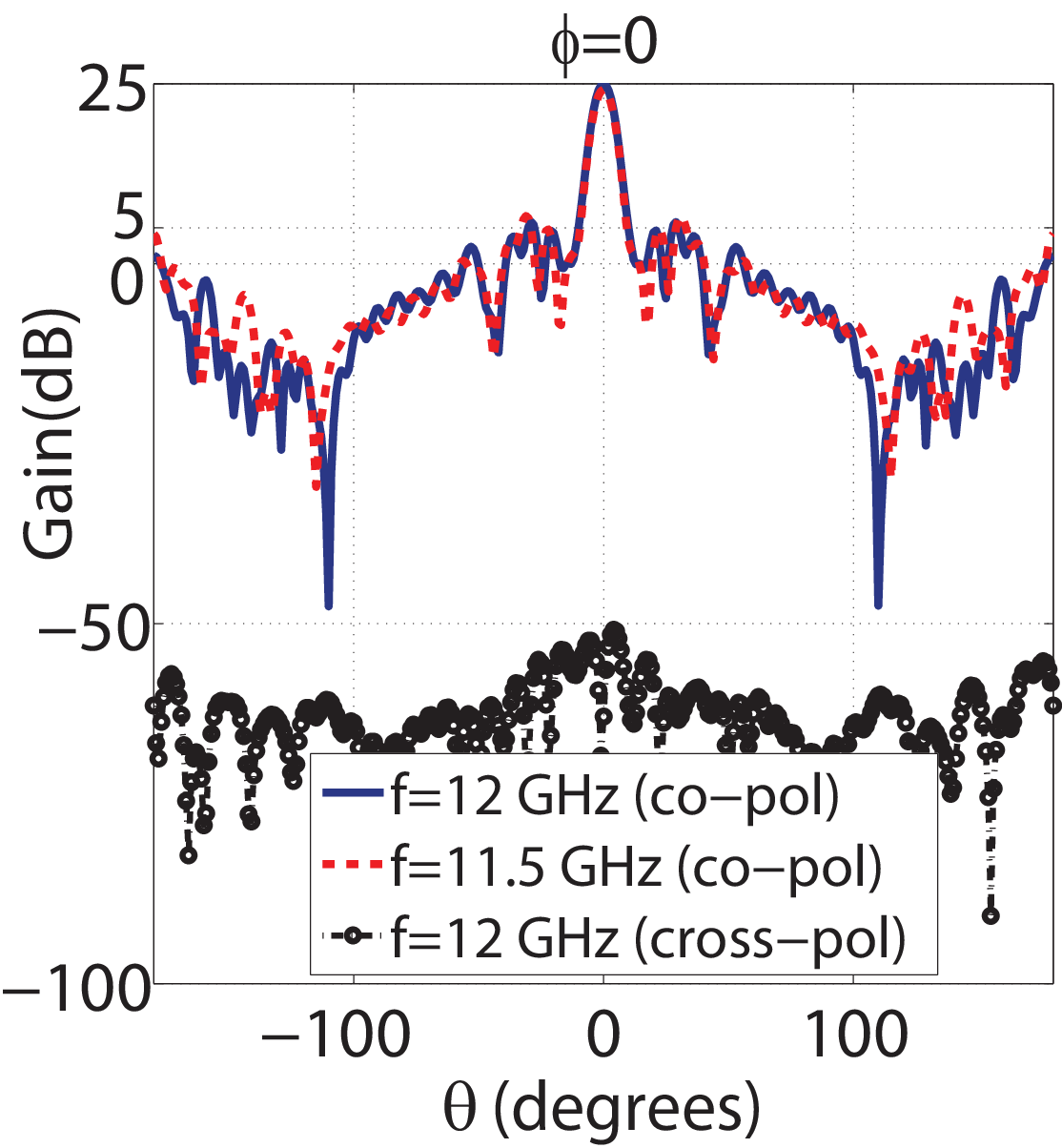}
		\caption{}
		\label{fig:fig10(a)}
	\end{subfigure}
	\begin{subfigure}[b]{0.32\textwidth}
		\includegraphics[width = \textwidth]{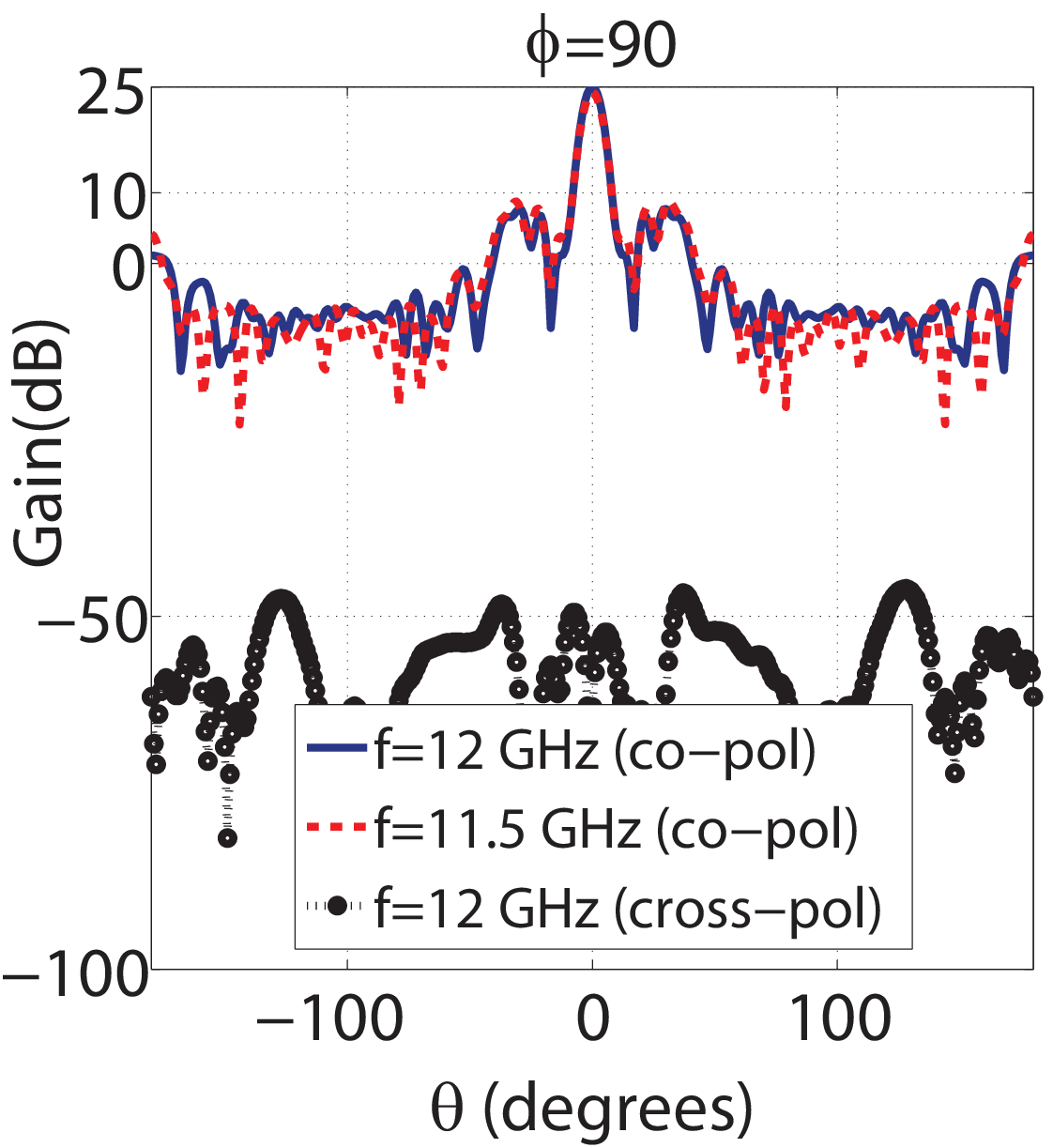}
		\caption{}
		\label{fig:fig10(b)}
	\end{subfigure}
	\begin{subfigure}[b]{0.332\textwidth}
		\includegraphics[width = \textwidth]{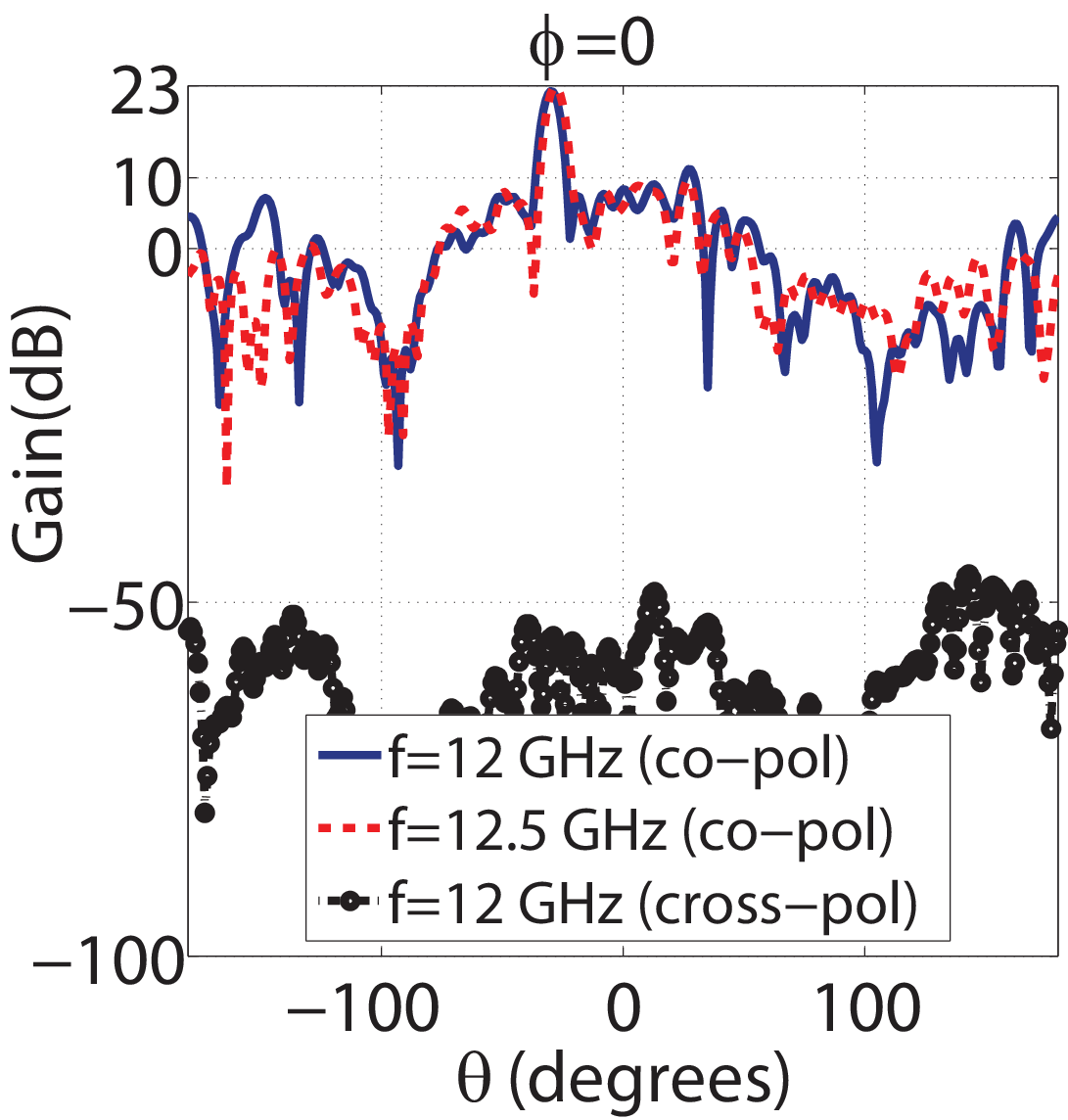}
		\caption{}
		\label{fig:fig10(c)}
	\end{subfigure}
\begin{subfigure}[b]{0.32\textwidth}
		\includegraphics[width = \textwidth]{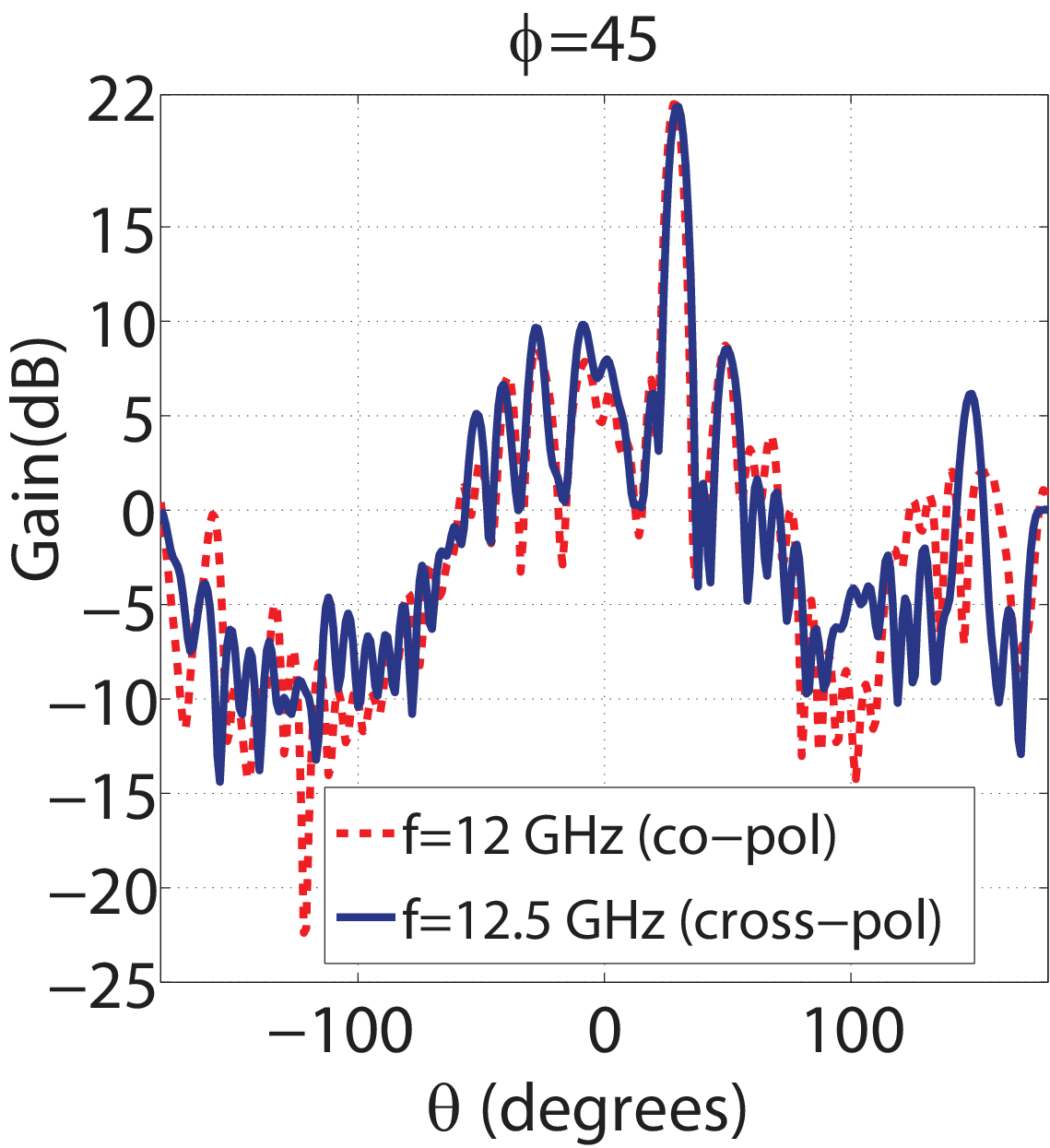}
		\caption{}
		\label{fig:fig10(d)}
	\end{subfigure}
	\begin{subfigure}[b]{0.322\textwidth}
		\includegraphics[width = \textwidth]{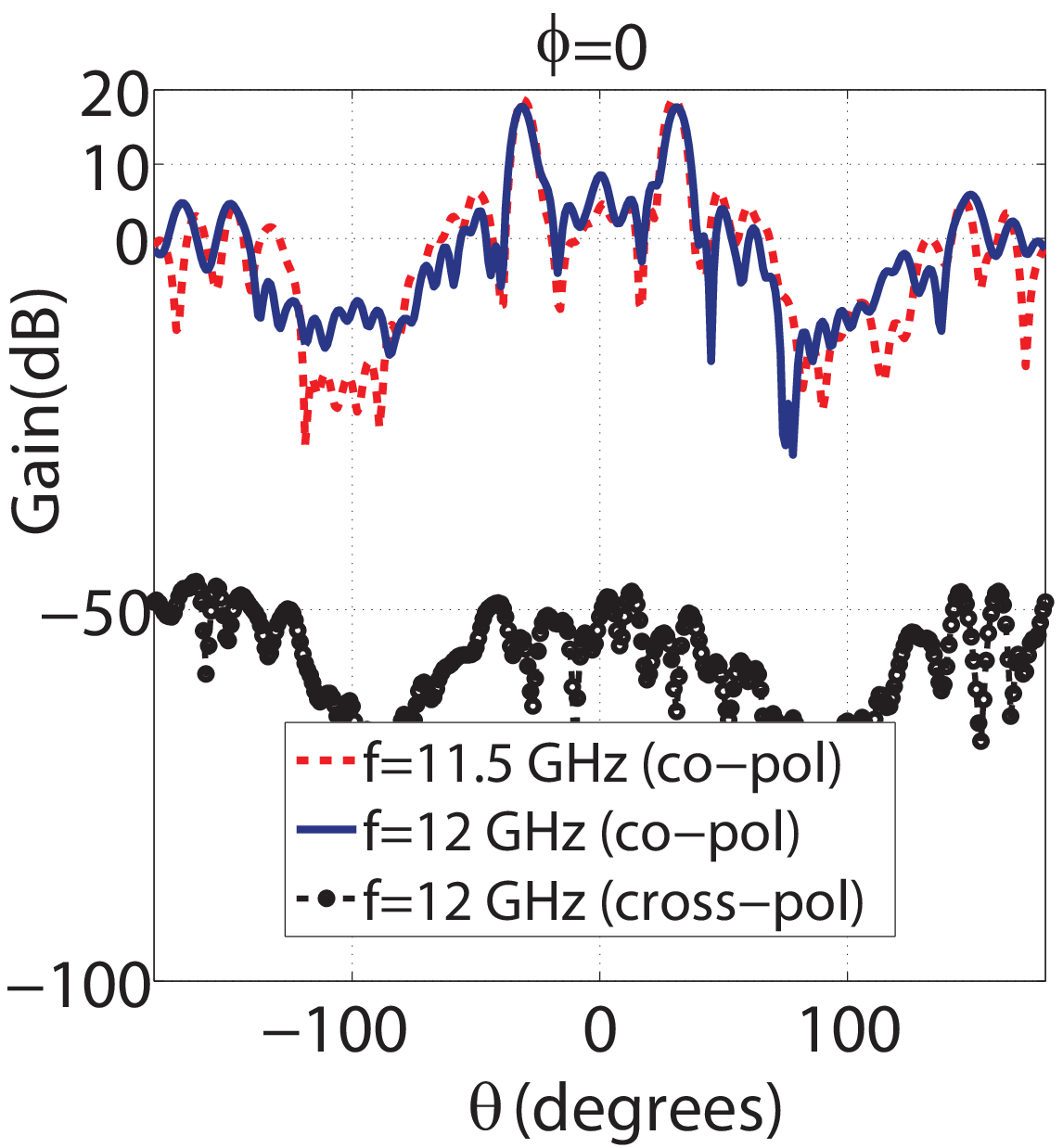}
		\caption{}
		\label{fig:fig10(e)}
	\end{subfigure}
\caption{Simulated radiation pattern of the holograms. (a) Example 1  ($\theta=0^\circ,\phi=0^\circ$) (b) Example 1 ($\theta=0^\circ,\phi=90^\circ$), (c) Example 2 ($\theta=30^\circ,\phi=0^\circ$), (d) Example 3 ($\theta=30^\circ,\phi=45^\circ$), and (e) Example 4 ($ \theta=30^\circ,\phi_1=0^\circ (\phi_2=180^\circ)$).}
\end{figure*}
In this section, several examples are provided to validate the solution proposed in section 4.
\subsection{Design Consideration}
The design frequency of all antennas is 12 GHz, making them a suitable candidate for X-band applications. Moreover, all antennas have a circular aperture with a radius of 13.2 cm. For the antenna feed, we have used the ATM-90-440-6 pyramidal horn antenna. The feed's radiation pattern is estimated by  $\cos^q(\theta)$, where q=5.2. At 12 GHz, the feed gain is 13.3 dB, and its E-plane and H-plane 10-dB beamwidths are ±$35.76^\circ$ and ±$38.5\circ$, respectively. F/D is chosen as 0.6589 to suppress unwanted backlobes and achieve suitable aperture efficiency, leading to edge tapering of -11.4 dB and the maximum achievable aperture efficiency of 64.49\%. Besides, in all examples, parameters of Eq.(5) are assigned as follows: X= - 0.000041, M=0.009680, A=0.01287 and B=- 0.01227. Note that A and B (see Fig.\ref{fig:fig8}) are the maximum and minimum of the available susceptance range, respectively.
\subsection{Simulation Results}
Four examples are provided to validate the proposed solution. In the first to third examples, holographic TAs are designed to radiate a pencil beam directed at ($0^\circ, 0^\circ$), ($30^\circ, 0^\circ$), and ($30^\circ, 45^\circ$), respectively. In the last example, a holographic TA is designed to radiate two pencil beams directed at($30^\circ, 0^\circ$) and ($30^\circ, 180^\circ$).  The susceptance distribution of the holograms and the CST simulation results are shown in Fig.\ref{fig:fig9(a)} up to Fig.\ref{fig:fig9(d)}, and Fig.\ref{fig:fig10(a)} up to Fig.\ref{fig:fig10(e)}, respectively. Expectantly, the main beam direction varies by changing the frequency. According to the examples, when the frequency shifts about 0.5 GHz from the center frequency (12 GHz), the main beam direction offsets about 2 degrees from $\theta_0$. This causes reductions in the operating bandwidth of the proposed TAs. However, the beam squint can be ignored for the main lobe direction near to ($0^\circ, 0^\circ$). Generally, time-consuming iterative algorithms like PSO are used to synthesize array antennas based on the only-phase approach, while the above examples show that using the holographic technique leads to good results without the need to use time-consuming iterative algorithms. Furthermore, the cross-pol level is less than -50 (dB) for every example.

\section{Measurement Results}
\begin{figure}[!t]
\centering
	\begin{subfigure}[b]{0.22\textwidth}
		\includegraphics[width = \textwidth]{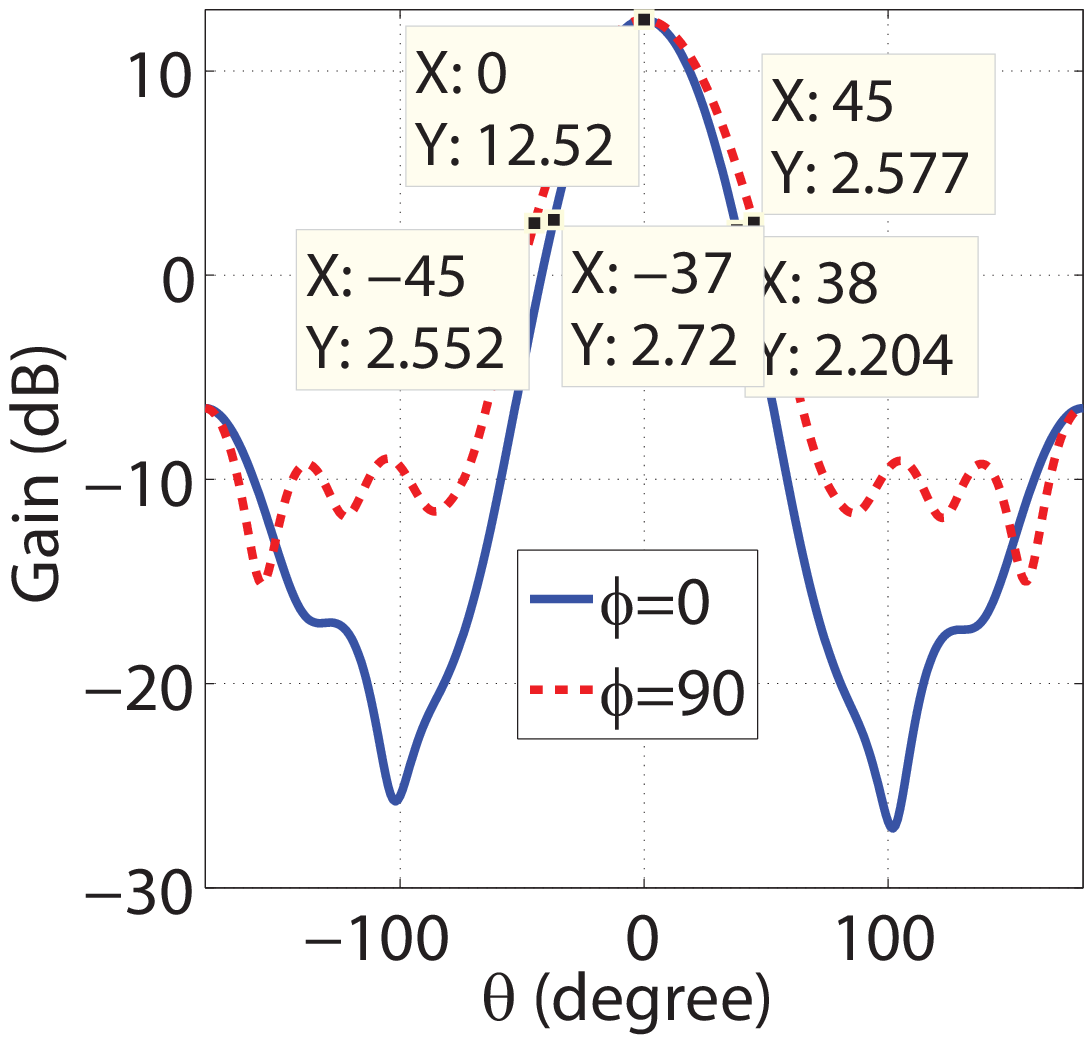}
		\caption{}
		\label{fig:fig11(a)}
	\end{subfigure}
	\begin{subfigure}[b]{0.22\textwidth}
		\includegraphics[width = \textwidth]{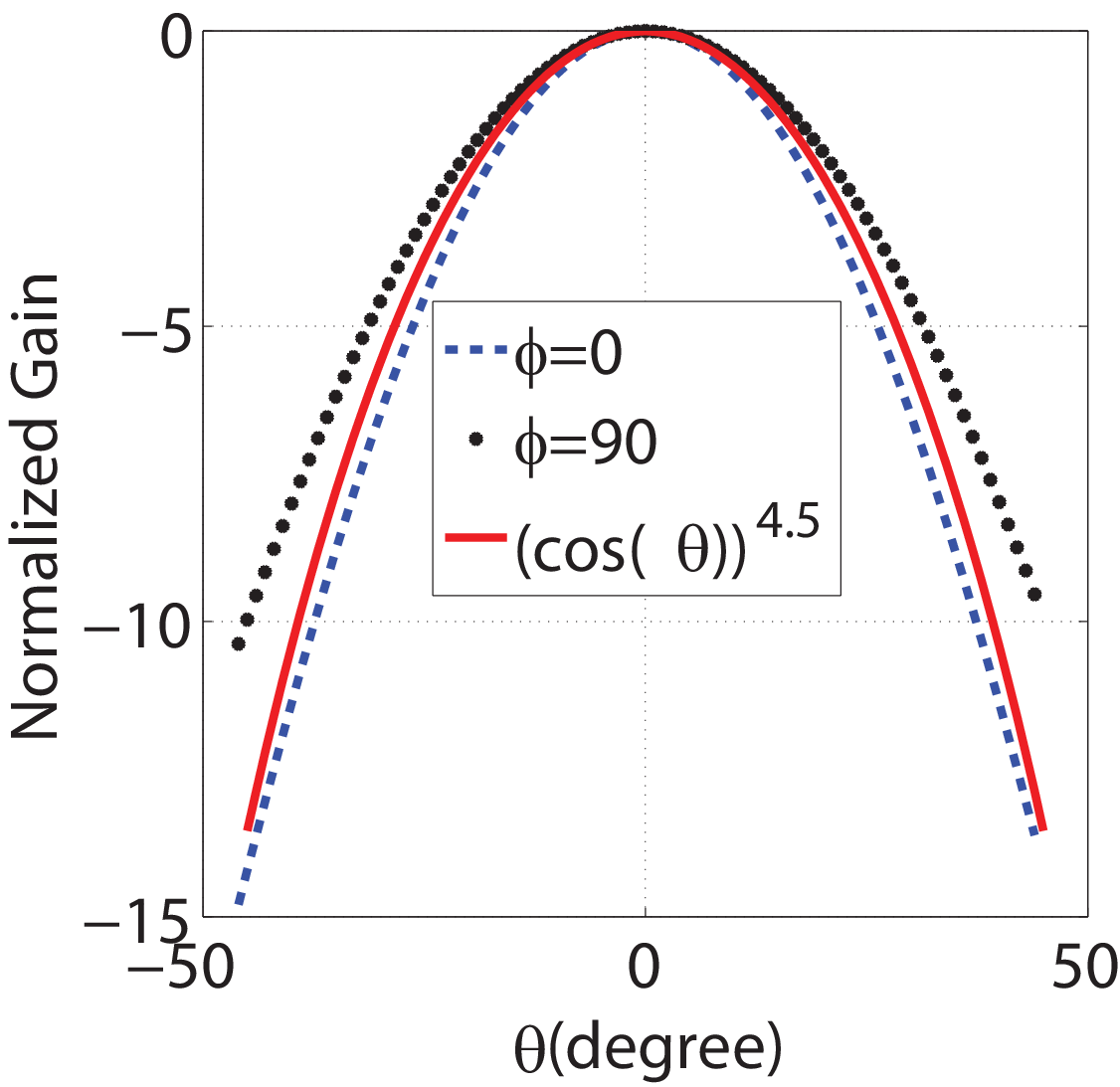}
		\caption{}
		\label{fig:fig11(b)}
	\end{subfigure}
	\begin{subfigure}[b]{0.22\textwidth}
		\includegraphics[width = \textwidth]{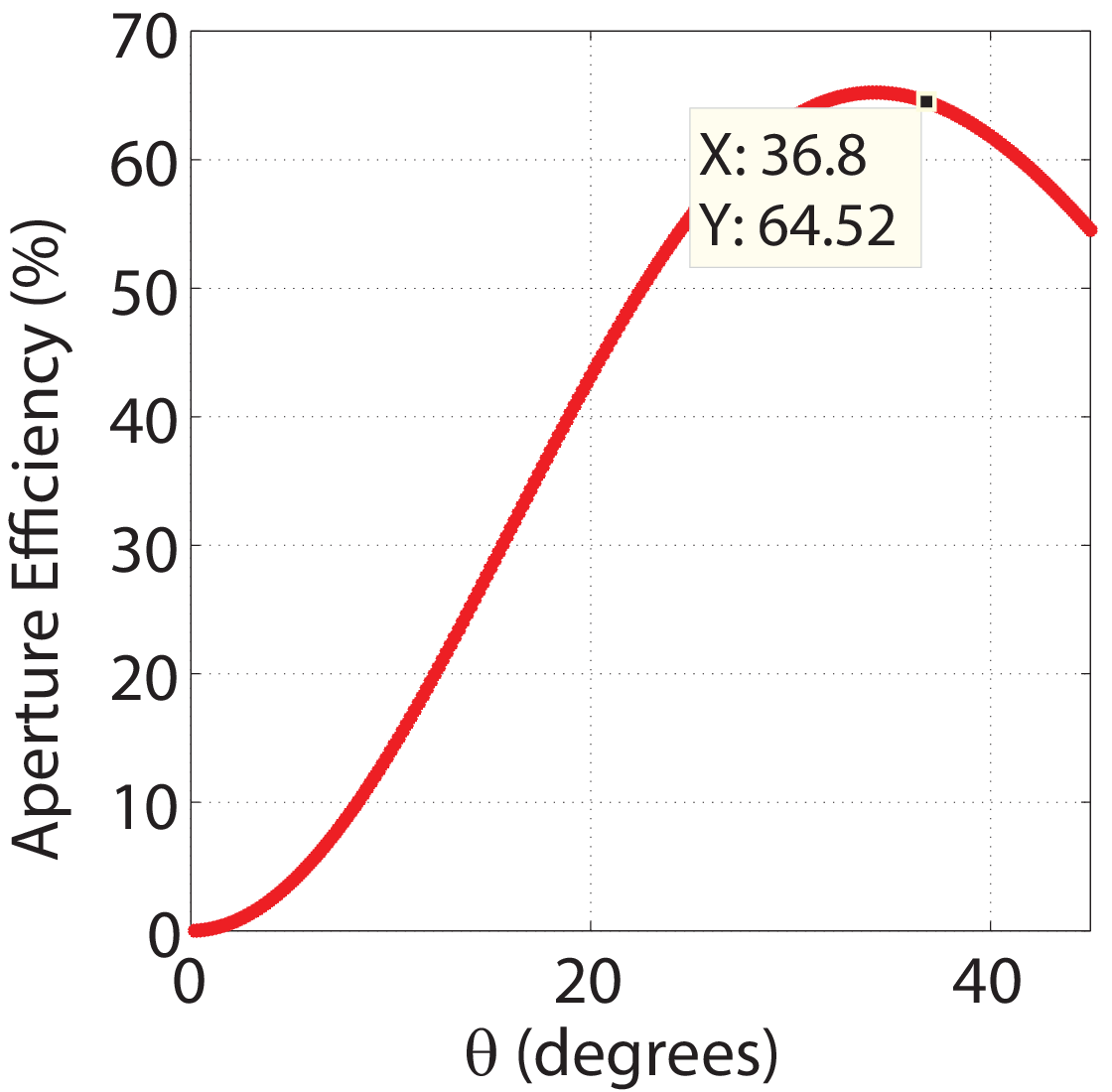}
		\caption{}
		\label{fig:fig11(c)}
	\end{subfigure}
	\begin{subfigure}[b]{0.22\textwidth}
		\includegraphics[width = \textwidth]{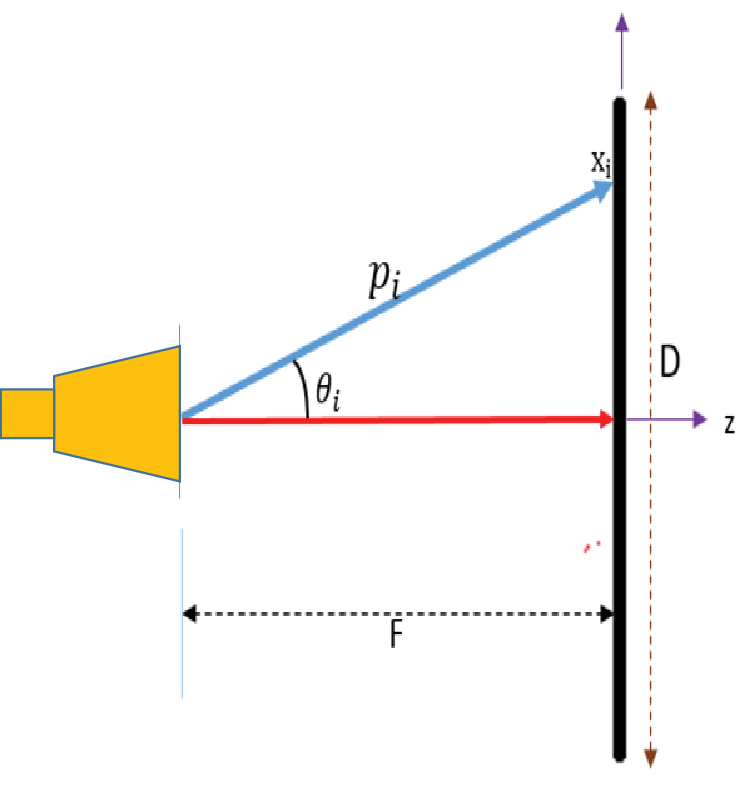}
		\caption{}
		\label{fig:fig11(d)}
	\end{subfigure}

\caption{(a) The E-plane and H-plane of the feed's radiation pattern, (b) Estimating the radiation pattern with $\cos^{4.5}(\theta)$, (c) the achievable aperture efficiency versus ($\theta$), and (d) the 2D schematic of the feed and transmitarary}
\end{figure}

As a proof of concept, a holographic TA using the unit cell shown in Fig.\ref{fig:fig6(a)} and Fig.\ref{fig:fig6(b)} is designed and manufactured. As the aperture efficiency of the square aperture antenna is lower than that of the circular one, a circular aperture holographic TA with a radius of 13.3 (cm) is fabricated. The antenna feed is a pyramidal horn whose aperture size is $40\times{26} ({cm}^2)$. It operates at X band and uses the WR90 waveguide. The feed's gain is 12.5 dB at 12 GHz, and its 10-dB beamwidths of E-plane and H-plane are ±$37.33^\circ$ and ±$45.07^\circ$, respectively, as shown in Fig.\ref{fig:fig11(a)}. The feed's radiation pattern is estimated with $\cos^{4.5}(\theta)$, according to Fig.\ref{fig:fig11(b)}. The achievable aperture efficiency is the product of the spillover ($\eta_s$) and illumination ($\eta_i$) efficiencies, which are defined in Eq.(12) and Eq.(13) as a function of ($\theta$) \cite{a15}. The achievable aperture efficiency is shown in Fig.\ref{fig:fig11(c)} versus ($\theta$). In addition, the relation between F/D and ($\theta$) is expressed in Eq.(14), according to Fig.\ref{fig:fig11(d)}. According to the 10-dB beamwidths of the feed radiation pattern, the F/D ratio is selected to be 0.659, leading to the maximum achievable aperture efficiency of 64.29\%, as seen in Fig.\ref{fig:fig11(c)}.

\begin{equation}
\eta_{i}=\frac{\bigl[\frac{1-(\cos^{q+1}(\theta))}{q+1}+\frac{1-(\cos^{q}(\theta))}{q})\bigr]^2}{2\tan^{2}(\theta)\frac{1-\cos^{2q+1}(\theta)}{2q+1}}\label{eq12}
\end{equation}
\begin{equation}
\eta_{s}=1-\cos^{2q+1}(\theta)\label{eq13}
\end{equation}
\begin{equation}
\frac{F}{D}=\frac{1}{2\tan(\theta)}\label{eq14}
\end{equation}
\\

\begin{figure*}[!t]
\centering
\begin{subfigure}[b]{0.30\textwidth}
		\includegraphics[width = \textwidth]{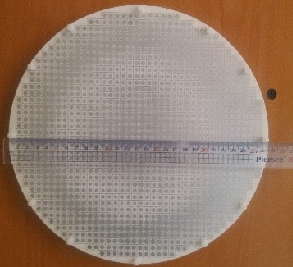}
		\caption{}
		\label{fig:fig12(a)}
	\end{subfigure}
	\begin{subfigure}[b]{0.36\textwidth}
		\includegraphics[width = \textwidth]{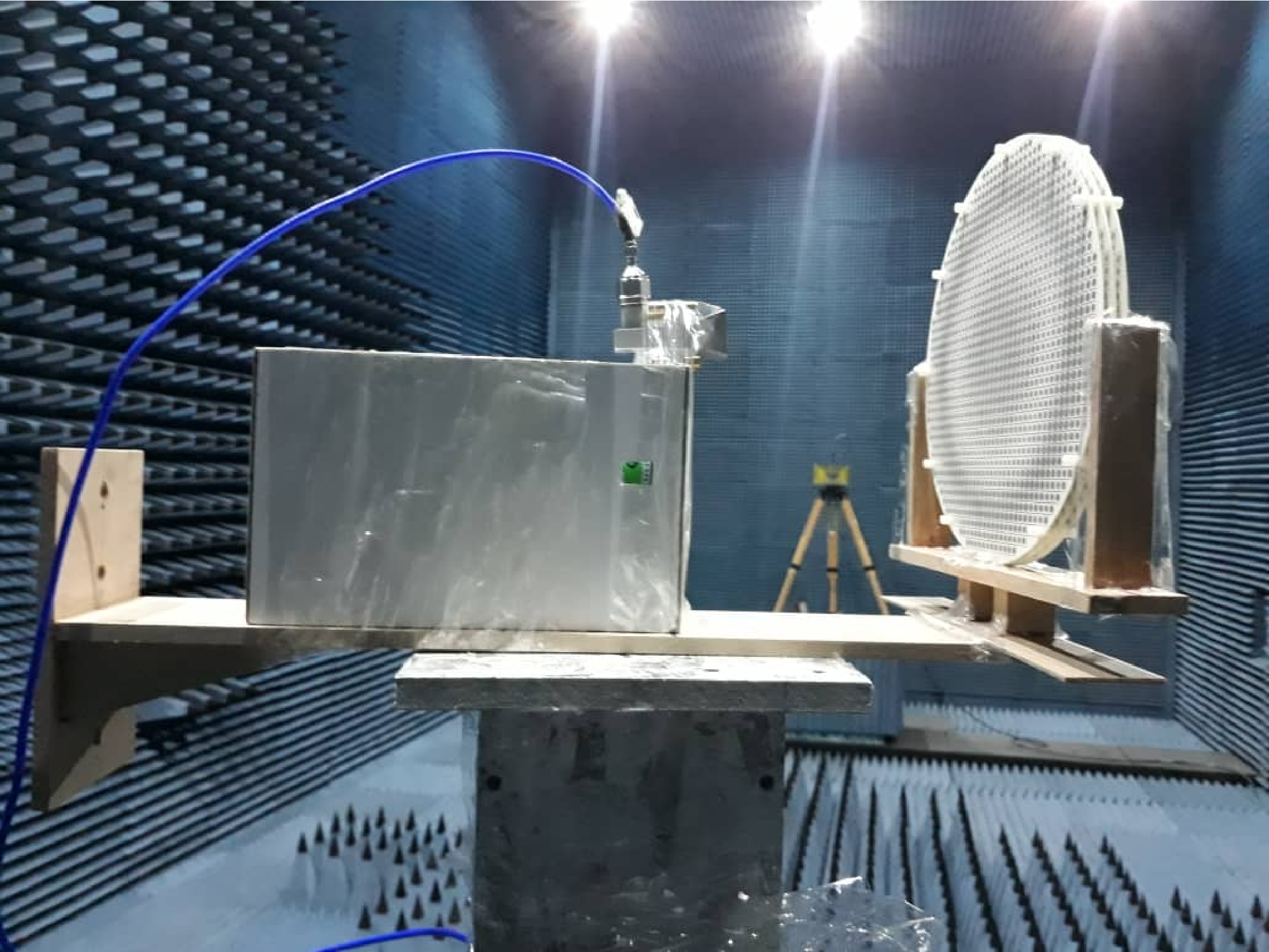}
		\caption{}
		\label{fig:fig12(b)}
	\end{subfigure}
\begin{subfigure}[b]{0.33\textwidth}
		\includegraphics[width = \textwidth]{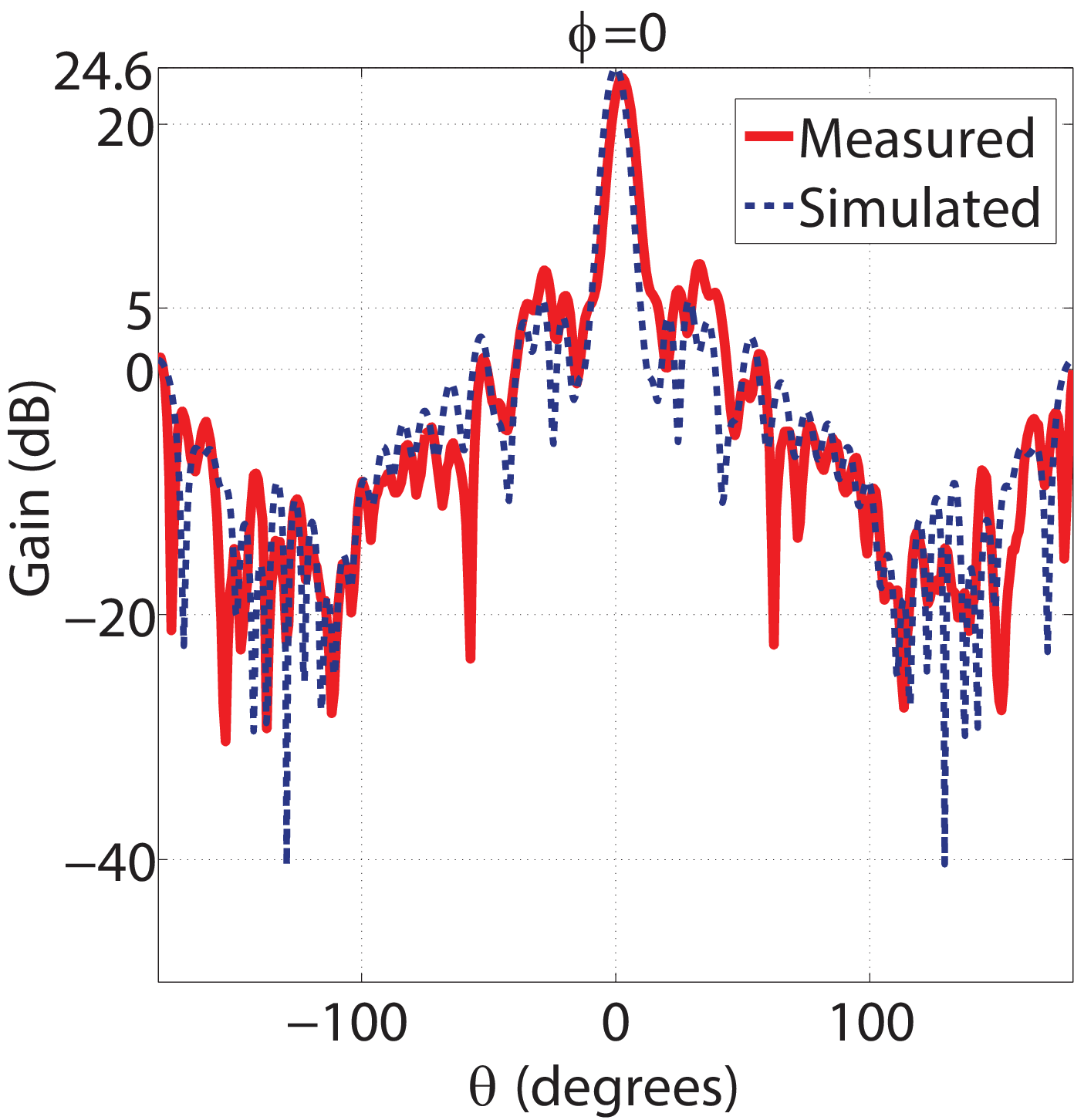}\quad
		\caption{}
		\label{fig:fig12(c)}
	\end{subfigure}
	\begin{subfigure}[b]{0.33\textwidth}
		\includegraphics[width = \textwidth]{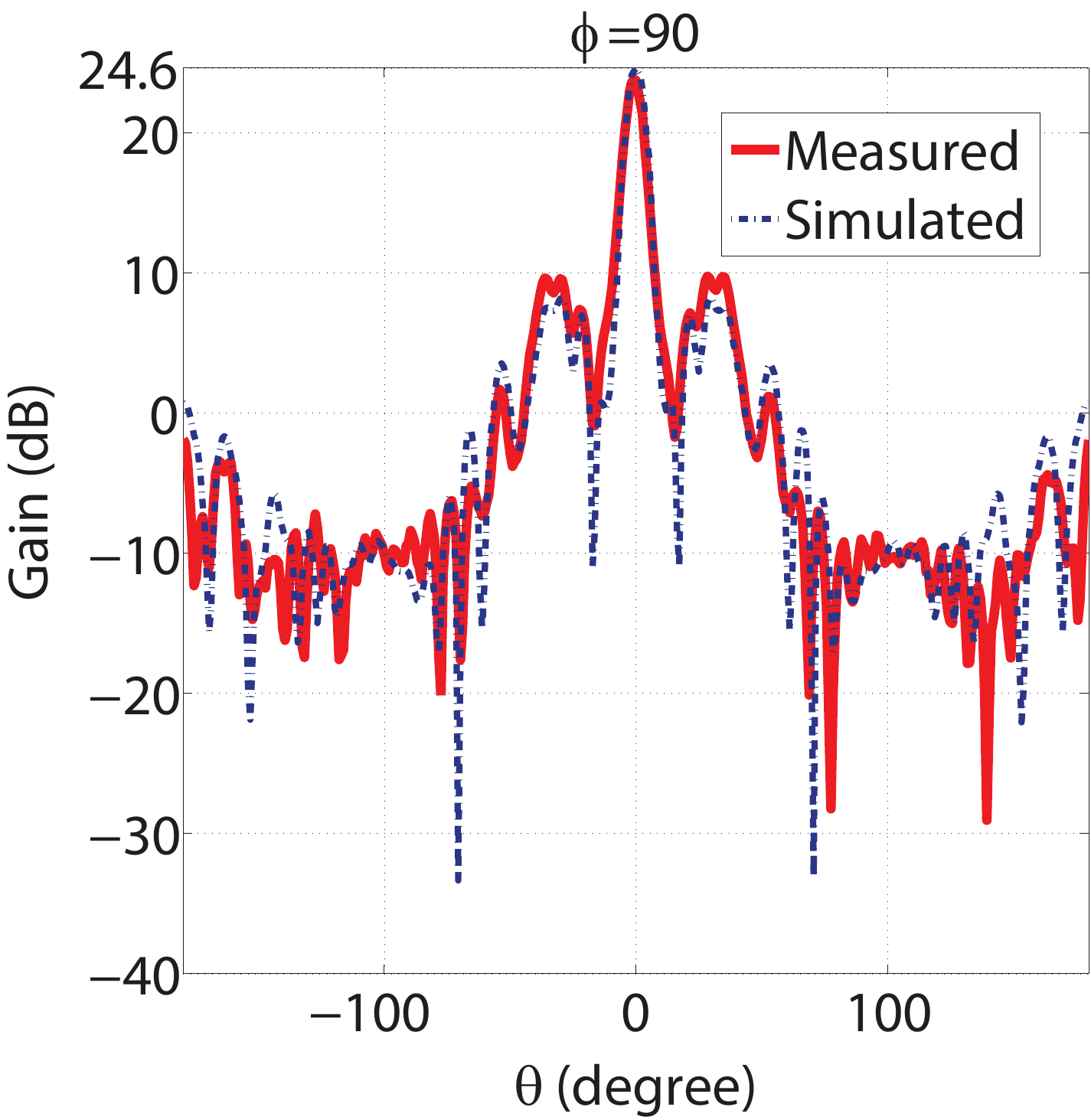}
		\caption{}
		\label{fig:fig12(d)}
	\end{subfigure}
	\begin{subfigure}[b]{0.54\textwidth}
		\includegraphics[width = \textwidth]{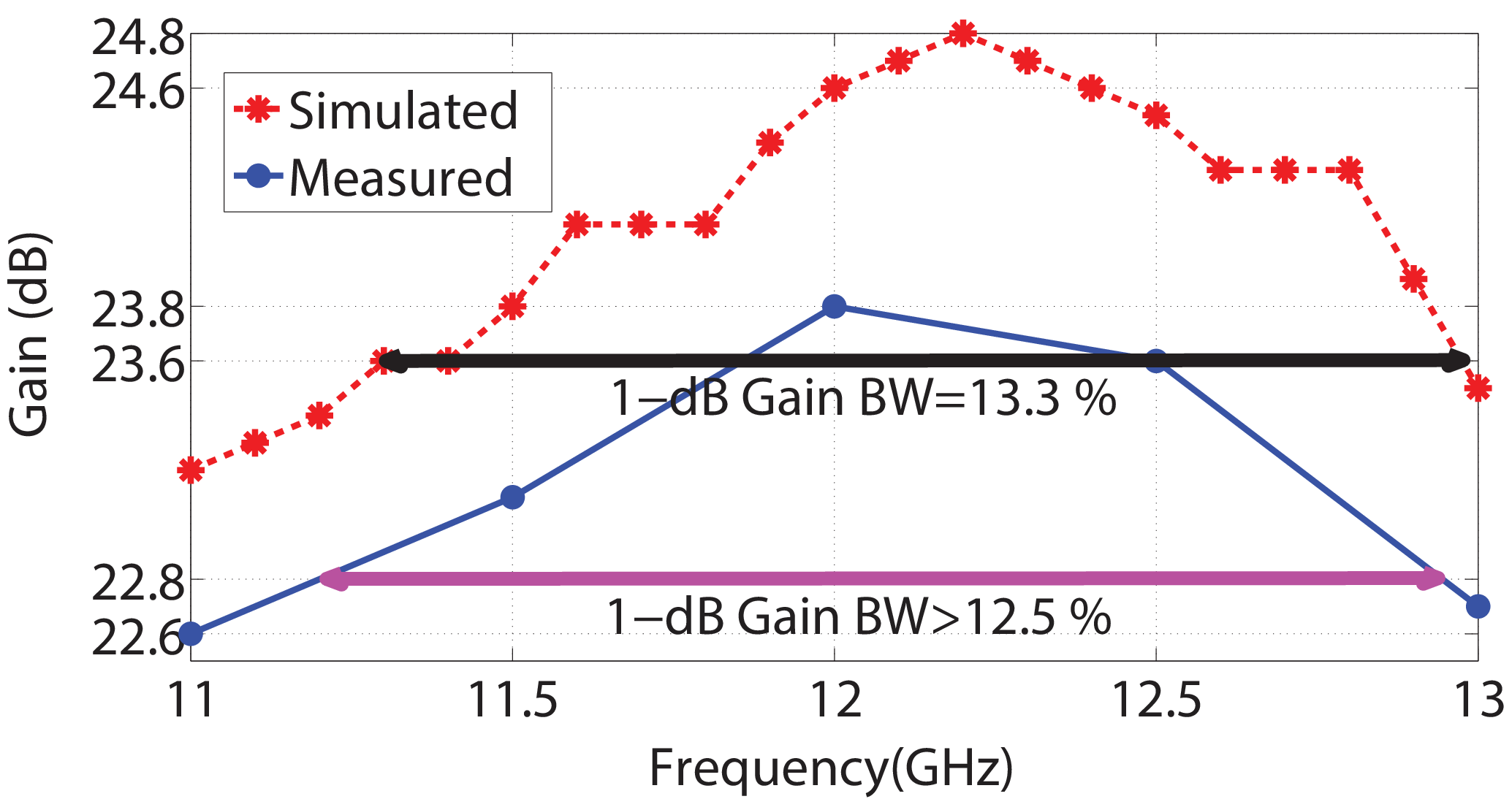}
		\caption{}
		\label{fig:fig12(e)}
	\end{subfigure}
\caption {(a) The manufactured transmitarray (b) The measurment setup (c) The measured and simulated gains at 12 GHz in yoz-plane,(d) The measured and simulated gains at 12 GHz in xoz-plane, and (e) The measured and simulated gains versus frequency.}
\end{figure*}

 Fig.\ref{fig:fig12(a)} and Fig.\ref{fig:fig12(b)} show the fabricated antenna and the measurement setup. The simulated and measured results are presented in  Fig.\ref{fig:fig12(c)} and Fig.\ref{fig:fig12(d)} at E and H planes. The figures show that the measured maximum sidelobe levelS in $\phi=0^\circ$ and $\phi=90^\circ$ planes are 7.59 dB and 8.59 dB, respectively. Furthermore, the measured and simulated gains in terms of frequency are presented in Fig.\ref{fig:fig12(e)}. The simulated gain increases from 23 dB at 11 GHz to 24.6 dB at 12 GHz and reaches its maximum value of 24.8 dB at 12.2 GHz. Then it decreases to 23.5 dB at 13 GHz. In addition, the measured gain increases from 22.6 dB at 11 GHz to 23.8 dB at 12 GHz. Finally, it decreases to 22.7 dB at 13 GHz. Because of the phase center misalignment, fabrication errors corresponding to the feed horn and the array antenna, approximations of the simulation model, and antenna setting errors in the anechoic chamber, the measured gain is up to 1 dB lower than the simulated gain. The figures show that, the simulated and measured 1-dB gain bandwidths are 13.3\% (11.3 GHz-12.9 GHz) and greater than 12.5\% (11.4 GHz-12.9 GHz). By substituting the measured (simulated)gain (G=23.8 (24.6) dB), aperture area ($A=\pi133^{2} (mm^{2})$), and wavelength ($\lambda_0=25 mm$) to Eq.(15) \cite{a16}, the simulated and measured aperture efficiencies are obtained equal to 25.8\% and 21.46\% at 12 GHz, respectively. Compared with the hologram proposed in Example 1 (section 6 "Hologram Design"), the simulated aperture efficiency is decreased from 28.3\% (Example 1) to 25.8\% due to differences in the dimensions of the applied horn feeds. Eq.(16) depicts the relation between the simulated aperture efficiency and the maximum achievable aperture efficiency \cite{a17}. The $\eta_{feed}$ factor equals one at the simulation stage, while the lossy $\eta_{cell}$ factor is less than one and reduces the maximum aperture efficiency value due to the following reasons:
  1- Using three layers of lossy dielectric (Rogers 4003C) in the structure of holograms reduces $\eta_{cell}$. 2- When the CST Studio software simulates the proposed unit cell, it considers an infinite number of identical unit cells (Periodic Boundary Condition), while the proposed holograms consist of a finite number of almost non-identical unit cells. This phenomenon reduces $\eta_{cell}$ because of the coupling between adjacent unit cells constructing a hologram which deteriorates the scattering parameter of the unit cell.

 \begin{equation}
AE(\%)=\frac{10\frac{G (dB)}{10}\lambda_{0}^{2}}{4\pi{A}}100\label{eq10}
\end{equation}
 \begin{equation}
AE(\%)=\eta_i\times\eta_s\times\eta_{feed}\times\eta_{cell}\label{eq11}
\end{equation}
\\

  \par The simulated radiation efficiency is 95.94\%. According to the law of power conservation, the input power($P_{in}$) the hologram receives from the feed horn equals the sum of the power radiated to the front hemisphere of the antenna aperture ($P_r$), the power radiated to the back of the antenna aperture($P_b$), and the power which is lost mainly by generating heat ($P_{loss}$). The radiation efficiency is the ratio of $P_r$ to $P_{in}$  \cite{a18}. If the unit cell periodicity decreases, the number of susceptance (reactance) samples of the mathematical hologram that can be realized by the unit cells increases, leading to more accurate hologram implementation, which results in decreasing $P_b$ and $P_{loss}$. According to the law of power conservation, when $P_b$ and $P_{loss}$ decrease, $P_r$ has to increase to make $P_{in}$ constant, which increases the radiation efficiency. Therefore, the achieved radiation efficiency (95.94\%) is mainly because of using subwavelength unit cells in constructing the holograms. Table 1 compares the proposed hologram with some existing works designed based on the only-phase technique. Regarding the 1-dB gain bandwidth, the proposed antenna performs better than Ref. 19 to 25, mainly because of considering the effects of almost all incident angles on the transmission susceptance calculations and using subwavelength unit cells, which is possible due to applying the holographic technique. In terms of the aperture efficiency, the performance of the hologram is better than Ref.19 and 25. However, it has worse performance than Ref.20 to 24. Ref.20 to 22 uses metal-only unit cells (getting rid of lossy dielectrics) to achieve high aperture efficiency. Since our in-house laboratory facilities are not good enough to manufacture metal-only TAs, available low-cost PCB technology is chosen to fabricate the proposed hologram, which reduces the aperture efficiency due to using lossy dielectric layers. Ref.23 and 24 achieve higher aperture efficiency than the proposed hologram because of using fewer dielectric layers. Although Ref. 25 uses two dielectric layers, it achieves lower aperture efficiency than the proposed hologram, as it does not optimize the focal distance to get the maximum aperture efficiency.

\begin{table*}[!t]
\centering
\caption{\label{tab:example}Achievement comparison of our work with some existing TAs.}
\begin{tabular}{|l|l|l|l|l|l|l|}
\hline
Ref &Freq (GHz) &Dielectric Layer  & Metal-only& Gain (dB)& Aperture Efficiency (\%)& 1 dB Gian Bandwidth (\%)\\
\hline
19 &12.5 &3 &No&18.9 &20.9 & 9.6  \\
20 &13.58 &4 &Yes&23.9 &55 &7.4 \\
21 &29.5 &3 &Yes&29.9 &50 & less than 11 \\
22 &11.45 &4 &Yes&24.26 &42 &4.2 \\
23 &9.7 &2 &No&23 &38 &5.7 \\
24 &11.3 &2 &No&28.9 &30 &9 \\
25 &10 &2 &No&22.7 &14.8 &9.6 \\
{This Work} &12 &3 &No&23.8 &21.46 & more than 12.5 \\

\hline
\end{tabular}
\end{table*}

\section{Conclusion}
 A detailed study on the design procedure of single-beam and dual-beam linearly polarized holographic transmitarrays, based on the susceptance (reactance) distribution is provided for the first time. Initially, the impedance surface of a metasurface is analyzed for both reflection and transmission modes. It is shown that generating only $180^\circ$ transmission phase range ($-90^\circ : 90^\circ$) is sufficient to cover all possible susceptance (reactance) values, which reduces the number of required dielectric layers and obsoletes four-layer transmitarrays. As the susceptance (reactance) values generated by an artificial impedance surface are different for the transmission and reflection modes, a new approach(different from reflectarray designs) based on the holographic technique is applied to record the amplitudes and phases of predetermined far-field radiations. Several holograms are designed to support the proposed solution. Finally, a holographic TA is manufactured and tested using available low-cost PCB technology as a proof of concept. The simulation and measurement results show that the antenna has good performance in terms of the 1-dB gain bandwidth (12.5\%), mainly due to considering almost all incident angles and using elements with dimensions equal to $0.24\lambda_0$, which becomes possible by applying the holographic technique. Besides, the antenna achieves 95.94\% simulated radiation efficiency and 21.46\% measured aperture efficiency.

\section*{Funding:}
This research did not receive any specific grant from funding agencies in the public, commercial, or not-for-profit sectors.

\section*{Author contributions statement}
This project was done under the guidance of Prof. Oraizi as the supervisor. Dr. Homayoon Oraizi suggested the project . Mr. Salehi, as the first author, proposed the method and performed the simulations. In doing so, he developed a Matlab code for synthesizing the holograms. All authors analyzed the measurement and simulation results. In addition, they reviewed and revised the manuscript.

\section*{Additional Information}
{\bf Competing Interests:} The authors declare no competing interests.

\end{document}